\documentclass[12pt]{iopart}
\usepackage{iopams}

\expandafter\let\csname equation*\endcsname\relax
\expandafter\let\csname endequation*\endcsname\relax

\usepackage{amsmath}
\usepackage{bm}
\usepackage{graphicx}
\usepackage{cite}
\usepackage[dvipsnames]{xcolor}
\usepackage[citecolor=RawSienna, colorlinks=true, linkcolor=MidnightBlue, urlcolor=MidnightBlue, linktocpage=true]{hyperref} % easy jump to the bibliography from a citation 
\usepackage{braket}

\begin{document}

\hspace*{\fill}FERMILAB-PUB-20-184-QIS

\title{Quantum Machine Learning in High Energy Physics}

\author{Wen Guan$^1$, Gabriel Perdue$^2$, Arthur Pesah$^3$, Maria Schuld$^4$, Koji Terashi$^5$, Sofia Vallecorsa$^6$, Jean-Roch Vlimant$^7$}
\address{$^1$ University of Wisconsin-Madison, Madison, WI, USA 53706}
\address{$^2$ Fermi National Accelerator Laboratory, Fermilab Quantum Institute, PO Box 500, Batavia, IL, USA 60510-0500}
\address{$^3$ Technical University of Denmark, DTU Compute, Lyngby, DK}
\address{$^4$ University of KwaZulu-Natal School of Chemistry and Physics, Durban, ZA 4000}
\address{$^5$ ICEPP, The University of Tokyo, 7-3-1 Hongo, Bunkyo-ku, Tokyo, JP 300-1153}
\address{$^6$ CERN, IT, 1, Esplanade des Particules, Geneva, CH 1211}
\address{$^7$ California Institute of Technology, PMA, Pasadena, CA, USA 91125-0002}

\ead{jvlimant@caltech.edu}
\vspace{10pt}
\begin{indented}
\item[]May 2020
\end{indented}

\begin{abstract}
Machine learning has been used in high energy physics for a long time, primarily at the analysis level with supervised classification.
%Deep learning has recently been adopted more widely, and at many levels of the data pipeline in high energy physics, thanks to success of deep learning met in other domains.
Quantum computing was postulated in the early 1980s as way to perform computations that would not be tractable with a classical computer.
With the advent of noisy intermediate-scale quantum computing devices, more quantum algorithms are being developed with the aim at exploiting the capacity of the hardware for machine learning applications. 
An interesting question is whether there are ways to apply quantum machine learning to High Energy Physics. This paper reviews the first generation of ideas that use quantum machine learning on problems in high energy physics and provide an outlook on future applications.
\end{abstract}

%\maketitle

\section{Introduction}

Particle physics is a branch of science aiming to understand the fundamental laws of nature by studying the most elementary components of matter and forces.
This can be done in controlled environments with particle accelerators such as the Large Hadron Collider (LHC), or in uncontrolled environments such as cataclysmic events in the cosmos.
The Standard Model of particle physics is the accomplishment of decades of theoretical work and experimentation. While it is an extremely successful effective theory, it does not allow the integration of gravity, and is known to have limitations.
Experimentation in particle physics requires large and complex datasets, which poses specific challenges in data processing and analysis.

Recently, machine learning has been played a significant role in the physical sciences. 
In particular, we are observing an increasing number of applications of deep learning to various problems in particle physics and astrophysics.
Beyond typical classical approaches \cite{Radovic:2018dip} (boosted decision tree, support vector machine, etc.), state-of-the-art deep learning techniques (convolutional neural networks, recurrent models, geometric deep learning, etc.) are being successfully deployed in a variety of tasks \cite{Albertsson:2018maf,Guest:2018yhq}.

The ambitious high luminosity LHC (HL-LHC) program in the next two decades and beyond will require enormous computing resources. It is interesting to ask whether new technologies such as quantum machine learning could possibly help overcome this computational challenge. 
The recent development of quantum computing platforms and simulators available for public experimentation has lead to a general acceleration of research on quantum algorithms and applications.
In particular, quantum algorithms have recently been proposed to tackle the computational challenges faced in particle physics data processing and analysis.
Beyond explicitly writing quantum algorithms for specific tasks~\cite{Shapoval:2019txi,Bapst:2019llh,Bauer:2019qxa,Zlokapa:2019tkn, Cormier:2019kcq}, quantum machine learning is a way to learn quantum algorithms to achieve a specific task, similarly to \textit{classical} machine learning.

This review paper of how quantum machine learning is used in high energy physics (HEP) is organized as follows.
An overview of the fields of quantum computing and quantum machine learning are first provided in Sections \ref{sec:QComp} and \ref{sec:QML}.
We review the applications of quantum machine learning algorithms for particle physics using quantum annealing in Sections \ref{sec:QAnneal} and quantum circuits in Section \ref{sec:QCircuit}.
We provide a field-wide view of unpublished work and upcoming results in Section \ref{sec:comingsoon}.
We conclude with discussions on the future of quantum machine learning applications in HEP in Section \ref{sec:discussion}.

\section{Quantum Computing}\label{sec:QComp}

More than three decades after Richard Feynman’s proposal of performing simulations using quantum phenomena \cite{feynman}, the first practical quantum computers are finally being built. 
The scope of calculations has significantly expanded, with a range of promising applications emerging, including optimization \cite{qaoa, brandao-sdp, childs-convex-optimization}, chemistry \cite{quantum-chemistry-1, quantum-chemistry-2}, machine learning \cite{qml-1, qml-2, qml-3}, particle physics~\cite{Shapoval:2019txi,Bapst:2019llh,Bauer:2019qxa,Zlokapa:2019tkn, Cormier:2019kcq}, nuclear physics \cite{linear-response, neutrino-nucleus, qc-atomic-nucleus} and quantum field theory \cite{preskill-qft-simulation, lattice-gauge-theory, subatomic-many-body, non-abelian-gauge-field}.

\subsection{Quantum circuit model}

Quantum computers were formally defined for the first time by David Deutsch in his 1985 seminal paper \cite{deutsch}, where he introduced the notion of a \textit{quantum Turing machine}, a universal quantum computer based on qubits and quantum circuits. 
In this paradigm, a typical algorithm consists of applying a finite number of quantum gates (unitary operations) to an initial quantum state, and measuring the expectation value of the final state in a given basis at the end.
Deutsch found a simple task that would require a quantum computer less steps to solve than all classical algorithms, thereby showing that quantum Turing machines are fundamentally different and can be more powerful than classical Turing machines.
Since then, many quantum algorithms with a lower computational complexity than all known classical algorithms have been discovered, the most well-known example being Shor's algorithm to factor integers exponentially faster than our best classical algorithm \cite{shor-algorithm}.
Other important algorithms include Grover's algorithm invented in 1996 to search an element in an unstructured database with a quadratic speed-up \cite{grover-algorithm}, and the Harrow-Hassidim-Lloyd (HHL) algorithm, invented in 2008 to solve linear systems of equations\cite{hhl-algorithm}.

However, all those algorithms require large-scale fault-tolerant quantum computers to be useful, while current and near-term quantum devices will be characterized by at least three major drawbacks:
\begin{enumerate}
    \item Noise: the \textit{coherence time} (lifetime) of a qubit and the \textit{fidelity} of each gate (accuracy of the computation) have increased significantly during the past years, but are still too low to use the devices for applications beyond small proof-of-principle experiments involving only a few qubits -- even if tricks like error-mitigation are used (see for example \cite{corcoles2019challenges}).
    %Quantum error-correction is a promising way to prevent noisy computations, but requires a certain noise threshold---not yet achieved---in order to be useful.
    \item Small number of qubits: current near-term quantum computers consist of between 5 and 100 qubits, which is not enough for traditional algorithms such as Shor's or Grover's to achieve a quantum advantage over classical algorithms. While steady improvements are made, increasing the number of qubits is not just a matter of scaling current solutions: Problems of connectivity, cross-talk, and the consistent quality of qubits require new engineering approaches for larger systems.
    \item Low connectivity: most current quantum devices have their qubits organized in a certain lattice, where only nearest-neighbors can interact. While it is theoretically possible to run any algorithm on a device with limited connectivity---by "swapping" quantum states from qubit to qubit---the quantum advantage of some algorithms can be lost during the process \cite{gosset-quantum-mean-value}.
\end{enumerate}

Therefore, a new class of algorithms, the so-called Near-term Intermediate-Scale Quantum (NISQ) algorithms \cite{Preskill_2018}, have started to emerge, with the goal of achieving a quantum advantage with those small noisy devices. One of the main classes of NISQ algorithms is based on the concept of \textit{variational circuits}: fixed-size circuits with variable parameters that can be optimized to solve a given task. They have shown promising results in quantum chemistry \cite{quantum-chemistry-1} and machine learning \cite{schuld18cc} and will be discussed in more detail in Section \ref{sec:QVC}.

\subsection{Quantum annealing}\label{sec:QA}
Another paradigm of quantum computing, called \textit{adiabatic quantum computing} (or \textit{quantum annealing}, QA) was introduced several years after the gate model described above \cite{nishimori-qa, mit-qa} and has been implemented by the company D-Wave. In theory, this paradigm is computationally equivalent to the circuit model and  Grover's algorithm can for instance be ported to quantum annealing \cite{adiabatic-grover}.
It is based on the continuous evolution of quantum states to approximate the solution of \textit{Quadratic Unconstrained Binary Optimization} (QUBO) problems, of the form:

\begin{align} \label{eqn:qubo}
    \min_{\bm{x}} E(\bm{x})=\sum_{i,j=1}^n J_{ij} x_i x_j + \sum_{i=1}^n h_i x_i
\end{align}
where $x_i \in \{0,1\}$ and $J_{ij}$ and $h_i$ are real numbers defining the problem. 

This general problem belongs to the complexity class NP-Hard, meaning that it can probably not be solved exactly in polynomial time even by a quantum computer\footnote{While a proof is still to be found, complexity theorists believe that quantum computers will not lead to exponential speed-ups for NP-Complete or NP-Hard problems}. Quantum annealing is a heuristic proposed to approximate the solution of a QUBO problem, or even solve it exactly when the input parameters $J_{ij}$ and $h_i$ have some particular structures \cite{adiabatic-grover}. 

More precisely, solving a QUBO instance is equivalent to finding the ground-state of the problem Hamiltonian
\begin{align}
    H_P=\sum_{i,j=1}^n J_{ij} \sigma_i^z \sigma_j^z + \sum_{i=1}^n h_i \sigma_i^z
\end{align}\label{equ:HP}
where $\sigma_i^z$ is the Z-Pauli matrix applied to the $i^{th}$ qubit. Quantum annealing consists of initializing the system in the ground-state of a simpler Hamiltonian, such as
\begin{align}
    H_I=\sum_{i=1}^n \sigma_i^x
\end{align}
and slowly evolving the system from $H_I$ to $H_P$ during a total time $T$, for instance by changing the Hamiltonian along the trajectory:
\begin{align}
    H(t)=\left(1-\frac{t}{T}\right) H_I + \frac{t}{T} H_P
\end{align}
The quantum adiabatic theorem tells us that if the transition between the two Hamiltonians is "slow enough", the system will stay in the ground-state during the whole trajectory, including at the end for our problem Hamiltonian. Measuring the final state will therefore give us the solution to our QUBO problem. The main caveat of this approach is that the maximum allowed speed of the evolution can fall rapidly with the system size (sometimes exponentially low), removing any potential advantage compared to classical algorithms. Knowing if a given problem (or class of problems) can take advantage of quantum annealing is an open research question, which is why research on quantum annealing applications has been driven largely by empirical studies.

Many optimization problems, including in machine learning, can be mapped to a QUBO instance, making quantum annealing an attractive platform for quantum machine learning, as developed in Section~\ref{sec:QAO}.

\section{Quantum Machine Learning}\label{sec:QML}

Quantum machine learning has evolved in recent years as a subdiscipline of quantum computing that investigates how quantum computers can be used for machine learning tasks -- in other words, how quantum computers can learn from data \cite{biamonte17, qml-3}. One can approach this question in three different ways, which reflect similar angles established in quantum computing: 
\begin{itemize}
\item the foundational approach that reformulates learning theory in a quantum setting \cite{arunachalam2017guest, ciliberto2020fast},
\item efforts to find quantum algorithms that speed up machine learning with regards to computational complexity measures \cite{lloyd14, rebentrost2014quantum, kerenedis16, ciliberto2018quantum},
\item a near-term perspective that develops new machine learning applications tailor-made for NISQ devices \cite{perdomo2018opportunities}
\end{itemize}

Currently, classical machine learning is a distinctively empirical discipline, pioneered by research conducted in industry. It is therefore not surprising that quantum machine learning research is also dominated by the near-term perspective, a fact reflected in the selection of papers discussed in this review.

The near-term perspective of quantum machine learning starts from the quantum devices available today and asks how they can be used to solve a machine learning problem. 
Circuit-based quantum computers have been predominantly used to \textit{compute the prediction} of a quantum machine learning model that can be trained classically \cite{farhi18, schuld18cc}, while quantum annealers have been proposed to \textit{optimize}  classical models \cite{benedetti16b, neven08bin}.  

\subsection{Quantum circuits as trainable models}\label{sec:QVC}

A machine learning model can often be written as a function $f(x,\theta)$ that depends on an input data point $x$ -- for example describing the pixels of an image or a vectorized text document -- as well as trainable parameters $\theta$. 
The result of the model, $f$, is interpreted as a prediction, e.g. revealing the label of $x$ in a classification task. 
For simplicity, we will here assume a scalar output.

We know from the basics of quantum mechanics that the result of a quantum circuit is a measurement with a probabilistic outcome -- for example, a qubit measured in state $\left| 0 \right>$ or $\left| 1 \right>$. 
However, the \textit{expectation value} of a quantum observable -- a central concept in quantum theory -- is a deterministic value. In simple terms, the expectation value is the weighted average of a measurement result. For example, after taking $1000$ measurements (``shots'') of a qubit, of which $900$ resulted in the outcome $\left| 1 \right>$ an estimate of the expectation of the qubit state would be $0.9$.
We can interpret this expectation as a prediction, and the quantum circuit is thereby serving as a \textit{quantum classifier} or \textit{quantum machine learning model}.

\begin{figure}[htb]
\centering
\includegraphics[scale=0.4]{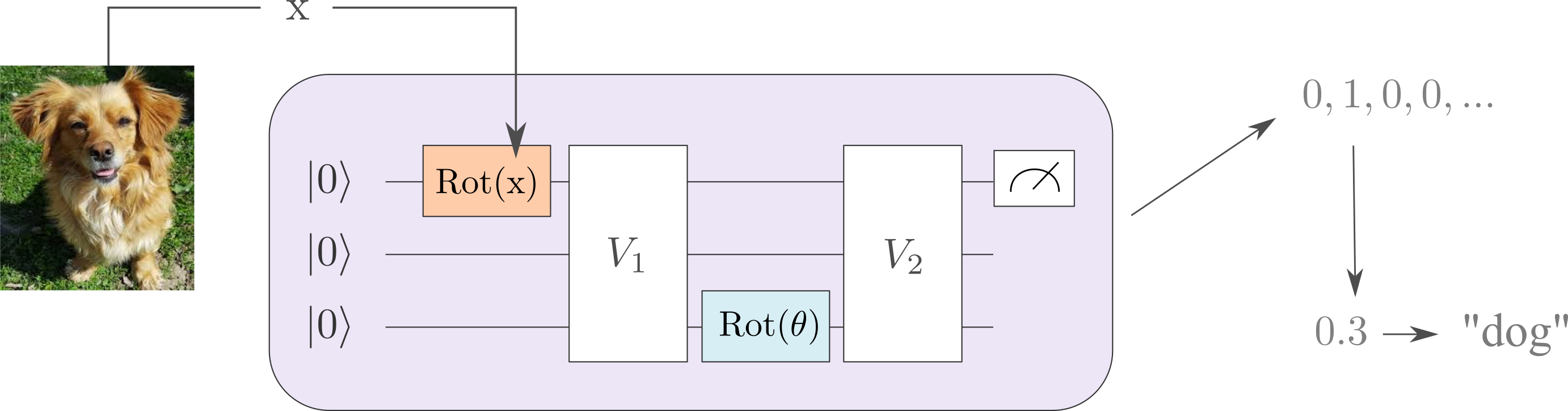}
\caption{Simplified example of a variational quantum circuit used for classification. A feature $x$ from the input data is loaded into the circuit by associating it with the angle of a rotation gate. The angle $\theta$ of another rotation gate is used as a variable parameter that can be trained to adjust the circuit. 
The three qubits are represented in the standard circuit notation as wires, and gates are represented by symbols acting on the wires. 
The unitaries $V_1$ and $V_2$ summarise arbitrary quantum operations applied to the qubits. 
The first qubit is measured in the end, and an expectation value is computed by averaging over measurement results. 
This expectation is interpreted as the prediction of a quantum model.}
\label{fig:VARCIRC}
\end{figure}

But how do we make the output of the quantum model depend on inputs $x$ and trainable parameters $\theta$? The central idea is to associate physical control parameters with the input features and individual parameters.
For instance, in most circuit-based quantum computers we have control over the rotation angle of qubits. 
Assuming for now that $x$ is a single scalar, we can therefore rotate one qubit by an angle of exactly $x$ to encode the input\footnote{Note that $x$ has to be rescaled to lie in the interval $[0, 2\pi]$ for the encoding to be unique.}.
Using the same strategy for a parameter $\theta$, considered to be a scalar as well for now, we can rotate another (or the same) qubit by an angle $\theta$. Physically, there is no difference in how the inputs and free parameters are treated, but there are profound conceptual differences; see for example \cite{lloyd2020quantum}.
These rotations can be performed as part of a larger quantum algorithm that consists of other gates, and which is described by an overall unitary $U(x, \theta)$ that depends on the input and parameter (see Figure~\ref{fig:VARCIRC}). 
The crux is that now the expectation value of the circuit with respect to an observable $M$ is formally given by 
$$f_{q}(x, \theta) =\left< 0 \left| U(x, \theta)^{\dagger} M  U(x, \theta) \right| 0\right>,$$ and can be interpreted as the prediction of $x$. In short, the quantum circuit is used as a machine learning model.

Of course, the heart of machine learning is to adapt a model to data. 
The circuit can be trained by adjusting the parameters $\theta$ by a classical optimization routine that minimizes a standard cost function comparing predictions with the correct target outputs, such as the mean square loss. 
Trainable circuits are also known as \textit{variational} or \textit{parametrized} circuits (or sometimes, a bit misleadingly, as \textit{quantum neural networks}), and were initially proposed in the context of quantum chemistry \cite{mcclean16}. 
The optimization can be performed by using the quantum computer to evaluate $f_{q}(x, \theta)$ at different values for $\theta$, and using a classical co-processor to find better candidates for the parameter with respect to the cost function, using either gradient-free or finite-difference based optimization methods.

Inspired from quantum control, quantum machine learning has recently developed an elaborate framework of gradient-based optimization \cite{mitarai2018quantum, schuld2019evaluating} that has already been implemented in powerful software frameworks \cite{bergholm2018pennylane, broughton2020tensorflow}, which may prove superior to gradient-free methods when quantum computers get bigger \cite{harrow2019low}. 
An essential result was to notice that in many cases used in practice, one can compute the analytic or exact gradient from $f_{q}(x, \theta+ s)$ and $f_{q}(x, \theta- s)$, where $s$ is a constant  which depends on the way that $\theta$ enters the quantum circuit -- in other words, which gate is used to encode the parameter.
While this is reminiscent of a finite-difference rule, the important fact is that $s$ is a macroscopic variable such as $\pi/2$, which makes estimating the two values by repeated measurements on a noisy device possible. 
Furthermore, the resulting gradient is not an approximation, but the true analytic gradient. 
The ability to compute gradients of variational circuits has potential consequences that reach far beyond quantum machine learning, since it makes quantum computing amenable for the paradigm of \textit{differentiable programming}.

Finally, it should be mentioned that there are many other ways that variational circuits are employed in quantum machine learning. 
For example, the genuinely probabilistic nature of quantum measurements suggests that variational circuits can be used as an ansatz for \textit{generative models}. 
In the generative mode, the result of a quantum measurement is interpreted as a sample of a probabilistic machine learning model that defines a probability distribution over the data that may depend on parameters \cite{liu2018differentiable, benedetti2019generative}. 
This has amongst other proposals led to \textit{quantum generative adversarial networks} \cite{lloyd2018quantum, dallaire2018quantum}.

\subsection{Quantum Annealers as Optimizers}\label{sec:QAO}

Quantum annealers represent a different approach to quantum machine learning. 
As natural optimizers, they outsource the \textit{training part} of machine learning to quantum computers, rather than the \textit{prediction part}. 
Since quantum annealers solve very specific optimization problems, more precisely QUBO problems (see Eq.~\ref{eqn:qubo}), the central challenge is to rephrase the loss function of a (quantum) machine learning problem in this format.

For example, an interesting and very early proposal \cite{neven08bin} recognized that the mean square loss of an ensemble of perceptrons -- the simple building blocks of neural nets -- can be written as a QUBO problem. 
A prerequisite is that the weights of the model have to be binary values - a condition that may even offer advantages for machine learning. The approach has been termed \textit{QBoost} and tested in one of the first commercial quantum annealers, the D-Wave machine, as early as in 2009 \cite{neven2009nips}. 
Other proposals to use the QUBO structure of quantum annealers for machine learning have been proposed for anomaly detection, in particular software verification and validation \cite{Pudenz:2011tue}.
 
Another, slightly different idea uses quantum annealers as samplers to support classical training of classical models \cite{benedetti16b}.
In the training of so-called Restricted Boltzmann Machines (RBMs), samples from a Gibbs distribution are required to find better candidates for the parameters in every step. 
The intimate connections between RBMs and Ising-type models in many-body physics (see also \cite{glasser2018neural} which reveals this connection through the language of tensor networks) suggest that quantum annealers, which are based on interacting spins, can produce samples from such Gibbs distribution. 
The details, especially when it comes to real hardware, are non-trivial, but successful quantum-assisted training has been demonstrated for small applications \cite{benedetti16b}. 
An important question raised as a result of this strategy was how samples from true quantum distributions, such as the Ising model with a transverse field, can be used to train \textit{quantum RBMs} \cite{amin2018quantum}.

\section{Quantum Annealing Applications}\label{sec:QAnneal}

For quantum annealers, the two most common approaches to machine learning involve mapping the problem into an optimization problem over the full dataset, and using the quantum device as a sampling engine to solve a difficult gradient calculation problem. 
In this section, we review papers that provide examples of these paradigms, we refer the reader to~\cite{QAML,alex2019quantum,Caldeira:2019lzf} for more in depth reading.

\subsection{Di-photon Event Classification}

The classification of collision events into signal or background categories is one of the main tasks in particle physics, and a frequent application for machine learning.
The Higgs boson, until its discovery in 2012 \cite{201230,20121}, was the missing piece of the standard model.
The authors of \cite{QAML} propose the use of quantum annealing to classify events between a Higgs decaying to a pair of photons and irreducible background events where two uncorrelated photons are produced.
To this end, eight high level features are measured from the di-photon system. 
With a view to using the method proposed in \cite{Pudenz:2011tue} --- so called \textit{quantum adiabatic machine learning (QAML)} --- a list of weak classifiers is computed from those eight features. 
Using the eight features and their products as input, $n=36$ weak classifiers ($c_i(x_\tau)$) are computed.
The weak classifiers assume values in the range $[-1,1]$ --- the signal being represented by positive values.
A strong classifier is then constructed from a binary linear combination of the weak classifiers (with parameter $w_i \in \{0,1\}$ for each weak classifier index $i$).

The parameters $w_i$ are then determined by the optimization of a carefully crafted QUBO
\begin{align}
E(\bm{w}) = \sum_{i,j=1}^{n=36} C_{ij} w_i w_j + \sum_{i=1}^{n=36} 2(\lambda - C_i) w_i
\end{align}
where $C_{ij} = \sum_\tau c_i(x_\tau) c_j(x_\tau) $ and $ C_{i} = \sum_\tau c_i(x_\tau) y_\tau $ are computed from the values of the weak classifiers in the training set and their category ($c_i(x_\tau)$ and $y_\tau$ respectively). $\lambda$ is a parameter penalizing solutions for too many weak classifiers participating. 
As described in Section~\ref{sec:QA}, the QUBO is transformed in a problem Hamiltonian $H_P$ (see Eq.~\ref{equ:HP}) with the change of variable $\sigma_i^z \leftarrow 2w_i -1$, and further embedded in a machine Hamiltonian to be solved on the device.
The set of parameters $w_i^*$ obtained through this optimization defines an optimal strong classifier as constructed above.

\begin{figure}[!thb]
   \begin{minipage}{0.48\textwidth}
     \centering
     \includegraphics[scale=0.25]{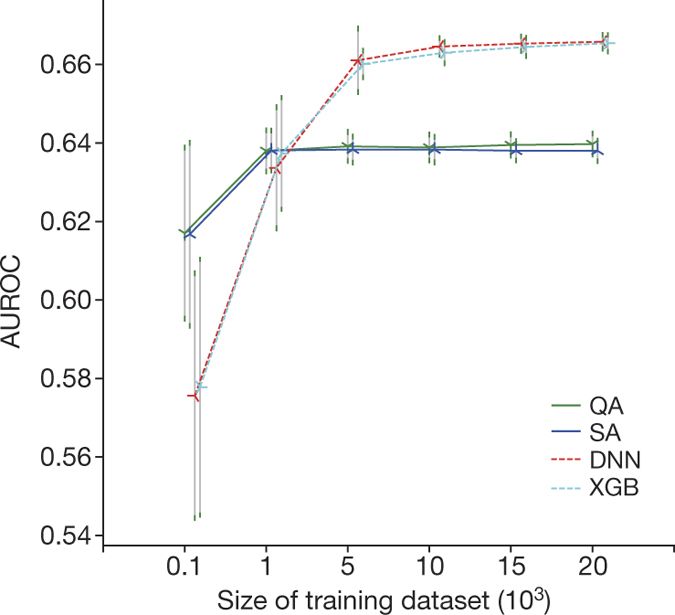}
     \caption{Area under the ROC curve (AUROC) of the strong classifier optimized on quantum annealer (QA) and simulated annealing (SA), together with the performance of boosted decision tree (BDT) and deep neural network (DNN) classifiers trained on the input features~\cite{QAML}.}\label{fig:QAML1}
   \end{minipage}\hfill
   \begin{minipage}{0.48\textwidth}
     \centering
     \includegraphics[scale=0.25]{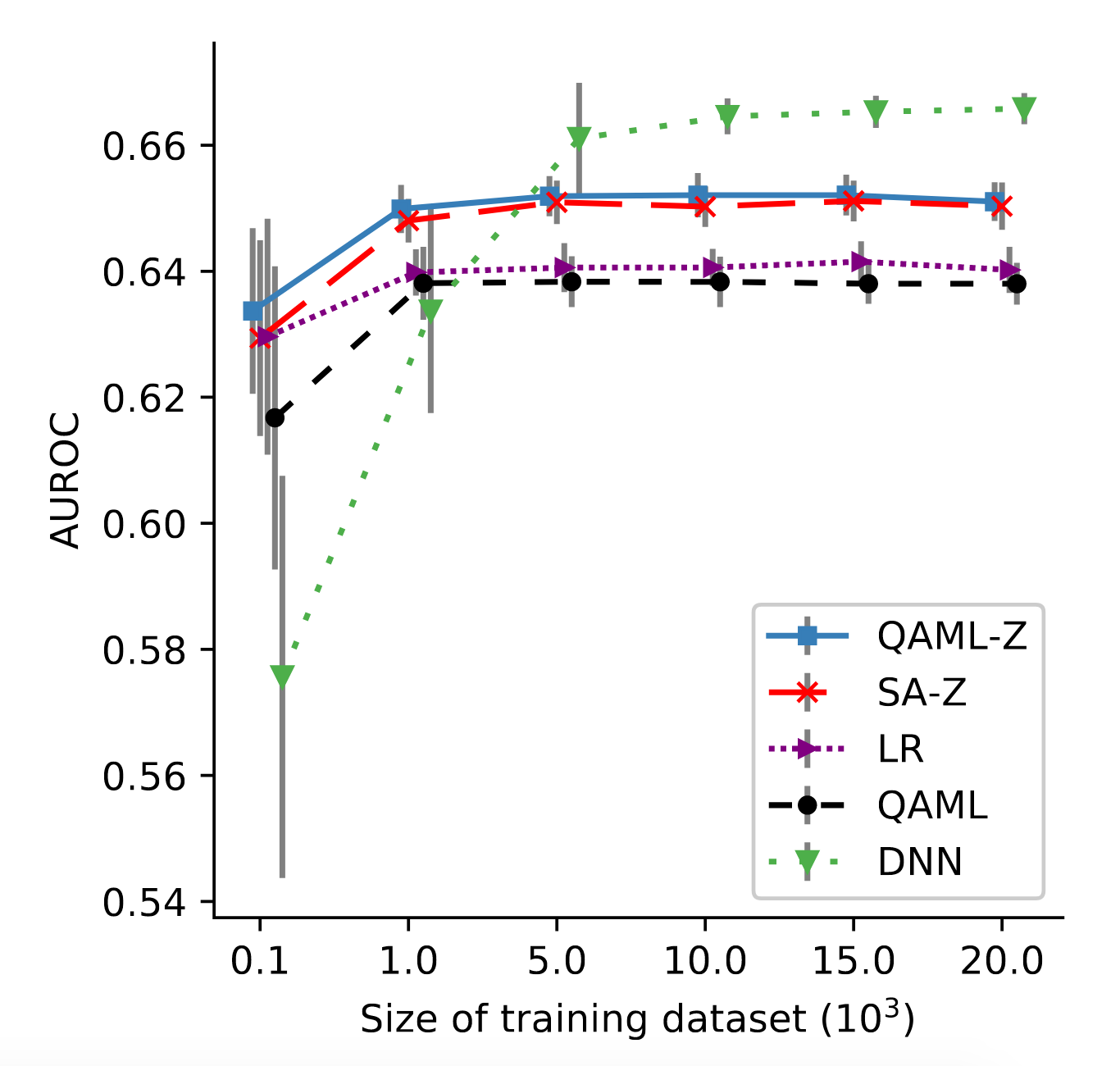}
    \caption{Area under the ROC curve (AUROC) of original QAML method, the continuous strong classifier optimized on quantum annealer (QAML-Z) and simulated annealing (SA-Z), together with the performance of a linear regression (LR) and deep neural network (DNN) classifiers~\cite{alex2019quantum}.}
\label{fig:QAML2}
   \end{minipage}
\end{figure}

The final performance of the strong classifier is compared with two classical machine learning methods: boosted decision tree (BDT) and deep neural network (DNN).
%To each of the lowest energy state found on QA corresponds a strong classifier, and the best performing one is selected.
The authors note that importance ranking can be obtained among the weak classifiers, by varying the parameter $\lambda$.
The optimization is both run on the D-Wave 2X quantum annealer system and performed with simulated annealing \cite{Kirkpatrick671,Katzgraber_2006} (SA) using variable fractions of the training dataset.
While SA is accurately finding the same ground truth found by QA, it is unable to reproduce the excited states measured with QA.
Therefore the inclusion of the excited states in the construction of the strong classifier with QA brings a slight, although not conclusive, difference in performance compared to the one derived with SA. 
SA and QA are typically \textit{on par}, and not providing obvious classification advantage over BDT and DNN (see Figure~\ref{fig:QAML1}), although a slight advantage with a small training dataset is noted.

In \cite{alex2019quantum}, the binary linear combination ($w_i \in \{0,1\}$) is extended to a continuous linear combination (denoted $\mu_i \in [0,1]$ to avoid confusion) by running the optimization in an iterative manner.
In order to take advantage of the continuous weights, additional weak classifiers, up to $N_w$ in total, are derived from the existing ones.
A new classifier is obtained from an existing one by shifting its value by a multiple of a given predefined shift, keeping the distribution clipped to the $[-1,1]$ interval.

The real parameters $\mu_i $ are obtained using the iterative rule 
\begin{equation}
    \mu_i(0) =  0 \;\;;\;\; \mu_i(t+1) = \mu_i(t) + \sigma_i^z(t) 2^{-(t+1)}
\end{equation}
where $\sigma_i^z(t)$ is the result of the optimization of the same Hamiltonian as in the binary case, evaluated under the change of variable 
\begin{equation}
    \sigma_i^z \leftarrow \mu_i(t) + \sigma_i^z(t) 2^{-t}
\end{equation}
We refer the reader to~\cite{alex2019quantum} for more details.
A bit flip heuristic is introduced between each iteration, with decreasing probability, as a regularization measure.
The authors note that there might be other such heuristic that could provide a better final accuracy.
The size of the problem Hamiltonian compared to the connectivity of the hardware is such that the authors prune cross-terms with low values and use a procedure provided by D-Wave to partially solve the optimization.
The proposed hybrid algorithm (so called QAML-Z) outperforms QAML while remaining without an accuracy advantage over classical approaches (see Figure \ref{fig:QAML2}).
Here again results obtained using simulated annealing and quantum annealing are \textit{on par}.
The scheme under which a discrete optimization is used iteratively as an approximation of continuous optimization using quantum annealers opens new directions for future algorithms.

\subsection{Classification in Cosmology with Quantum Restricted Boltzmann Machine}

Quantum annealers do not provide identical answers every time they go through an annealing cycle. 
For some applications it would be ideal if, for example, they always returned the lowest energy configuration, but instead they produce a distribution of states.
In principle, these states are Boltzmann-distributed with a characteristic temperature related to the physical device temperature.
In practice, the actual distribution of states deviates from a Boltzmann distribution (on the D-Wave 2000Q, for example, it is ``colder'' and tends to skew towards lower than expected energies).
However, with some post-processing the sample distribution may be converted into a Boltzmann distribution.
It may be also anecdotally observed that while the sampled distribution is not Boltzmann-distributed, simply applying the parameter update equations derived under the assumption of sampling from a Boltzmann distribution (see below, Equations \ref{eq:rbmupdateone} through \ref{eq:rbmupdatethree}) will generally allow the model to converge anyway \cite{Hinton2002,adachi2015application}.

Taken together, these observations mean that quantum annealers may also be used as sampling engines to fuel certain classes of machine learning algorithms.
Restricted Boltzmann Machines (RBMs) map well to modern quantum annealers for this purpose. 
They feature a bipartite connectivity graph that scales well in embedding algorithms as compared to a fully connected graph.
The tunable couplings between qubits function as graph connection weights and the annealing process naturally samples from the graph configurations with clamped or unclamped values for the visible nodes in the graph as needed by the application.

RBMs are fundamentally generative models that approximate a target distribution over an array of \textit{visible} binary variables ($\vec{v}$) as the marginal distribution of a bipartite graph that connects to a different set of \textit{hidden} binary variables ($\vec{h}$).
The distribution is described by
\begin{equation}
p(\vec{v},\vec{h}) \propto \exp(-v^{T}W h + b^{T}v + c^{T}h) \label{eqn:rbmprobdistribution}
\end{equation}
for some parameter (bias) vectors $\vec{b}$, $\vec{c}$, and a connections weight matrix $W$.

RBMs are trained by maximizing the log-likelihood of a data distribution by updating the bias and weights parameters.
With the loss ($L$) defined as the negative log-likelihood , the derivatives for the model parameters are
\begin{equation}
\label{eq:rbmupdateone}
\frac{\partial L}{\partial b^i} = \langle v^i \rangle_{\mathrm{data}} - \langle v^i \rangle_{\mathrm{model}}
\end{equation}
\begin{equation}
\label{eq:rbmupdatetwo}
\frac{\partial L}{\partial c^i} = \langle h^i \rangle_{\mathrm{data}} - \langle h^i \rangle_{\mathrm{model}}
\end{equation}
\begin{equation}
\label{eq:rbmupdatethree}
\frac{\partial L}{\partial W_i^j} = \langle v^i h^j \rangle_{\mathrm{data}} - \langle v^i h^j \rangle_{\mathrm{model}}
\end{equation}
These derivatives form a gradient for use in gradient descent for adjustments to $\vec{b}$, $\vec{c}$, and $W$.
The expectations are computed over the data (the training set) with clamped values and over the model with unclamped values. 
These steps are also referred to as the positive and negative phases.
See \cite{Adachi2015} for a particularly clear explanation.

While computing the expectations over the data is easy, computing the expectations over the model is costly, as that scales like $2^{\min{\left(n_v,n_h\right)}}$, with $n_v$ and $n_h$ equal to the number of visible and hidden units, respectively.
There are a number of mitigation strategies to avoid this difficult computation, all discussed in \cite{Caldeira:2019lzf}.
Of particular relevance here, the expectations for a given set of model parameters using unclamped variables on a D-Wave can be computed, where each computed configuration is sample from the machine's output distribution.
For small graphs this approach is impractical but it may eventually offer some computational advantage for very large graphs.

In practice, the distribution of states returned by the D-Wave 2000Q is not Boltzmann distributed, and significant post-processing is required to achieve a Boltzmann distribution.
As observed in \cite{Caldeira:2019lzf}, the D-Wave offers essentially no sampling advantage over random string initial states if using only Boltzmann distributions for the optimization.
However, it has been observed that RBMs may be optimized with imperfect gradients \cite{Hinton2002}.
Therefore, it is possible to greatly reduce the amount of required post-processing and still train effective models.

For the task of galaxy morphology classification, in \cite{Caldeira:2019lzf} it was observed that RBMs, regardless of the training methods, were less effective than gradient boosted trees (likely the best classical algorithm for structured data like the dimensionality reduced galaxy images).
Additionally, the best classical methods for discriminative training outperformed the quantum, generative training.
However, regardless of training strategy, RBMs offered a performance advantage for very small datasets that gradient boosted trees and logistic regression tended to badly overfit. 
Furthermore, early in the small dataset training runs, the quantum generative training outperformed the classical discriminative training.

\section{Quantum Circuit Applications}\label{sec:QCircuit}

As introduced in Section~\ref{sec:QVC}, circuits with varying parameters can be optimized to perform a specific task, e.g. classification.
The parameters of these circuits can be determined with gradient-based optimization method.
The following papers are following this approach for HEP specific classification tasks.
We refer the reader to~\cite{CHEP2019.Cenk,Chan:2019zwk,terashi2020event} for more in-depth reading.

\subsection{Quantum Graph Neural Networks for particle track reconstruction}

Quantum computers promise to greatly speed-up search in large parameter spaces. 
Charge particle tracking --- \textit{tracking} in short --- is the task of associating sparse detector measurements (a.k.a "hits") to the particle trajectory they belong to.
Tracking is the cornerstone of event reconstruction in particle physics.
Because of their ability to evaluate a very large number of states simultaneously, they may play an important role in the future of track reconstruction in particle physics experiments. 
Reconstructing particle trajectories with high accuracy will be one of the major challenges in the HL-LHC experiments \cite{hl-lhc}. 
\begin{figure}[b]
\centering
\includegraphics[scale=0.3,angle=270]{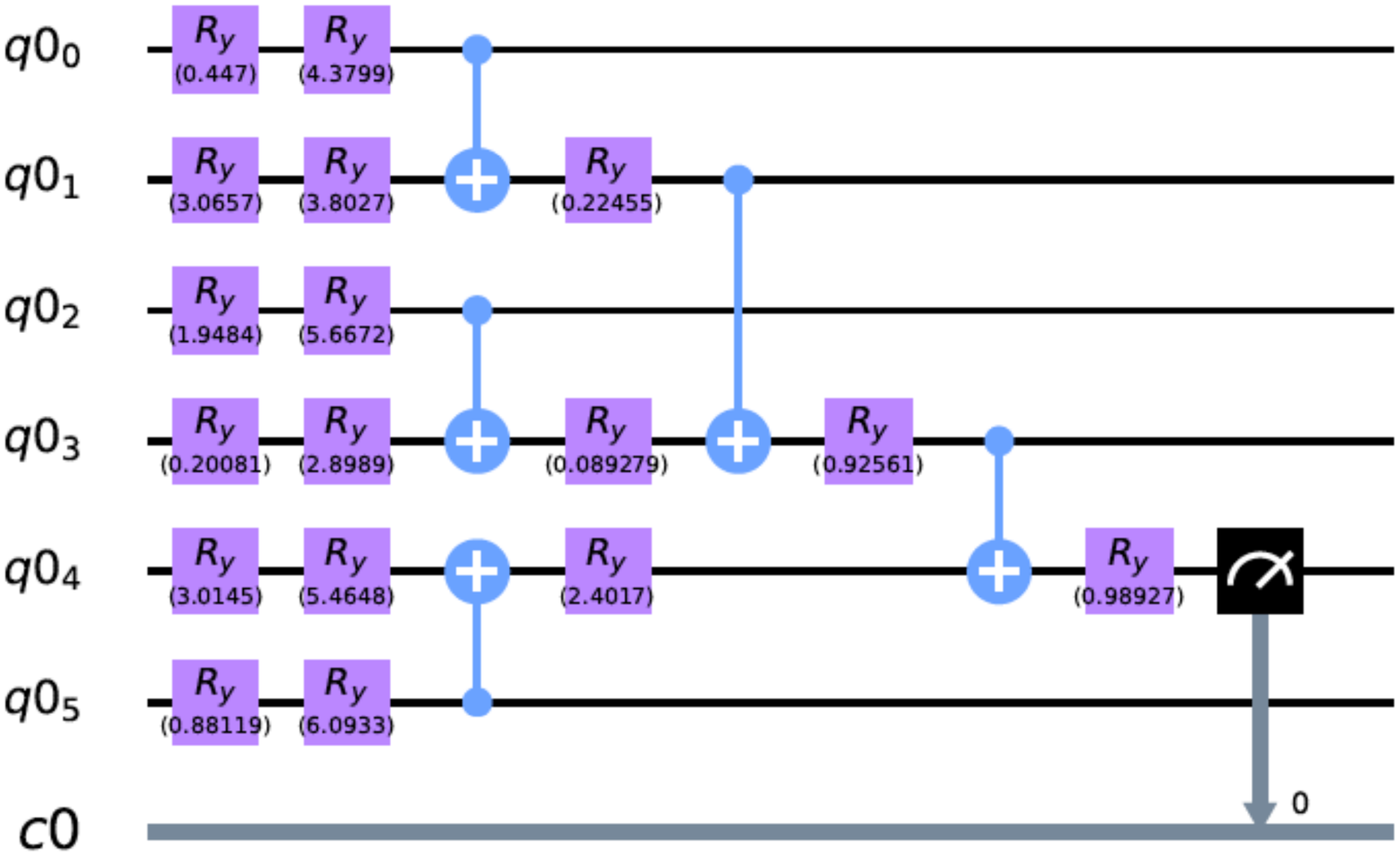}
\caption{The Quantum Edge Network implemented as a Tree Tensor Network, a hierarchical quantum classifier \cite{TTN}. The architecture uses $R_y$ rotation gates and CNOT gates. A single output qubit is measured. \cite{CHEP2019.Cenk} }
\label{fig:QGNN2}
\end{figure}

Increase in the expected number of simultaneous collisions and the high detector occupancy will make tracking  extremely demanding in terms of computing resources. 
State-of-the-art algorithms rely, today, on a Kalman filter-based approach: they are robust and provide good physics performance, however they are expected to scale worse than quadratically with the increasing number of simultaneous collisions \cite{hl-lhc}. 
The high energy physics community is investigating several possibilities to speed up this process \cite{Gumpert_2017,ATL-PHYS-PUB-2019-041,refId0} including deep learning-based techniques. 
For instance, introducing an image-based interpretation of the detector data and using convolutional neural networks can lead to high-accuracy results \cite{HEPtrk}. 
At the same time, a representation based on space-points arranged in connected graphs could have an advantage given high dimensionality and sparsity of the tracking data. 
The HEPtrkX  project  \cite{HEPtrk} followed this approach and successfully developed a set of Graph Neural Networks (GNNs) to perform hits and segments classification.
In this approach, graphs of connected hits are built, features of the graph nodes and edges are computed and, finally, relevant hit connections are predicted.
The dataset, designed for the TrackML challenge \cite{Amrouche_2019} contains precise locations of hits, and the corresponding particles. 
The classical GNN architecture consists of three networks organised in cascade: an \textit{input network} encodes the hits information as node features, an \textit{edge network} outputs edge features, using the start and end nodes, and a \textit{node network}, that calculate hidden nodes features taking into account all connected nodes on the previous and next layers. 
The edge and node networks are applied iteratively after the input network (see~\cite{farrell2018novel} for more details).
The work in \cite{CHEP2019.Cenk} represents an exploratory look at this GNN architecture from a quantum computing perspective: it re-implements the input, edge and node networks as quantum circuits.

\begin{figure}[tb]
\centering
    \begin{tabular}{c c}
\includegraphics[scale=0.6, trim={5cm 10cm 4.5cm 10cm}]{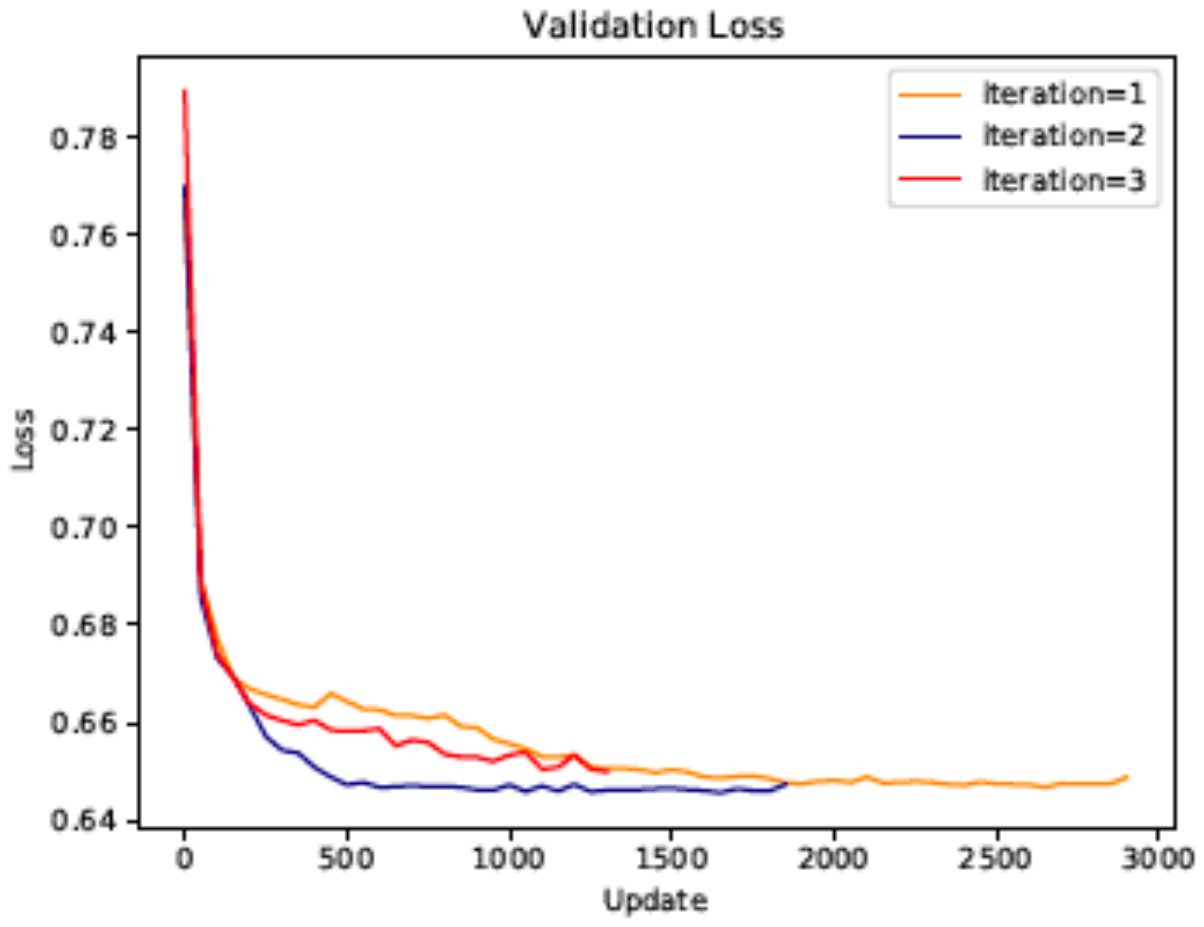}&
\includegraphics[scale=0.6, trim={5cm 10cm 4.5cm 10cm}]{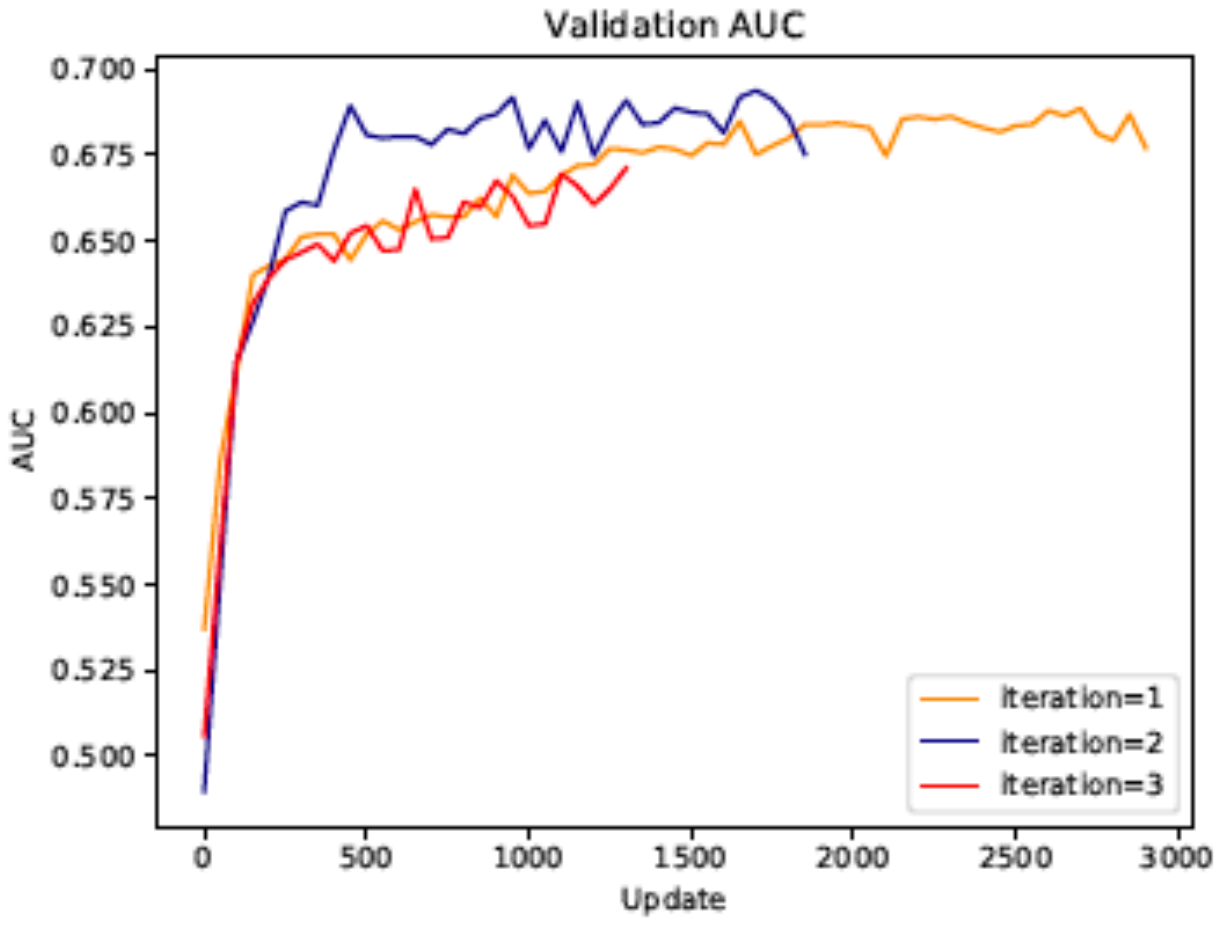} \\
    \end{tabular}
\caption{QGNN performance. The validation loss (on the left) decreases smoothly. Consistently, the validation accuracy (on the right) increases with the number of iterations. Results are shown for two epochs corresponding to 2900 steps (1 epoch = 1450 updates)~\cite{CHEP2019.Cenk} }
\label{fig:QGNN1}
\end{figure}

In particular, the edge and node networks are implemented as tree tensor networks (TTN) --- hierarchical quantum classifiers originally designed to represent quantum many body states described as high-order tensors \cite{TTN}.
%A classical input network is used to increase the dimension of the data. 
The data points are encoded (see Figure~\ref{fig:QGNN2}) as parameters of $R_y$ rotation gates
\begin{equation}
R_y (\theta) \left| 0\right> = \cos (\theta/2)\left| 0\right> + \sin (\theta/2)\left| 1\right>
\end{equation}
The TTN network consists of $R_y$ rotations and CNOT gates (see Figure ~\ref{fig:QGNN2}) and its output is the measurement from a single qubit.
The TTN has 11 parameters which are the angles of rotations in Y direction on the Bloch sphere. 
These parameters are optimized using the ADAM optimiser and a binary cross entropy loss function  using Pennylane \cite{bergholm2018pennylane} and Tensorflow \cite{tensorflow2015-whitepaper}.
The model is trained on 1450 subgraphs extracted from the TrackML dataset.

Although preliminary, the obtained performance  (see Figure~\ref{fig:QGNN1}) is promising: the validation losses decrease smoothly and the accuracy increases with the number of iterations. 
At convergence, the accuracy value is still lower than for the classical case.
This is, however, expected as the number of hidden features, and iteration are reduced compared to the GNN, because of computation issues.

\subsection{Classification Using Variational Quantum Circuits}

\begin{figure}[b!]
   \begin{minipage}{0.48\textwidth}
     \centering
     \includegraphics[scale=0.5]{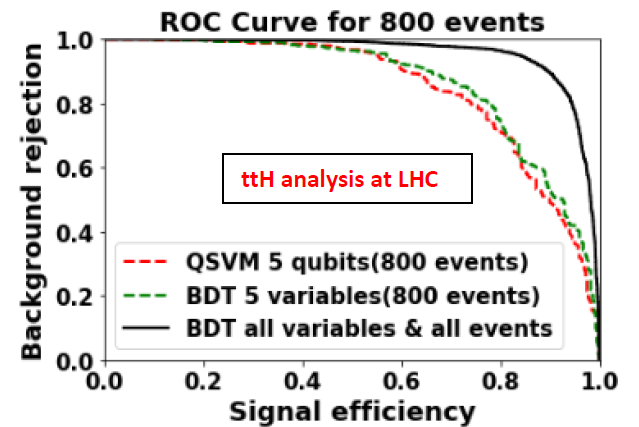}
     \caption{ROC curve of VQML and BDT methods. With 800 events and 5 qubits, the  VQML  have  obtained  very  close  performance  tothe  one  obtained  using  the  classical  machine  learning  method  BDT~\cite{Chan:2019zwk}. }\label{fig_qsvmroc}
   \end{minipage}\hfill
   \begin{minipage}{0.48\textwidth}
     \centering
     \includegraphics[scale=0.5]{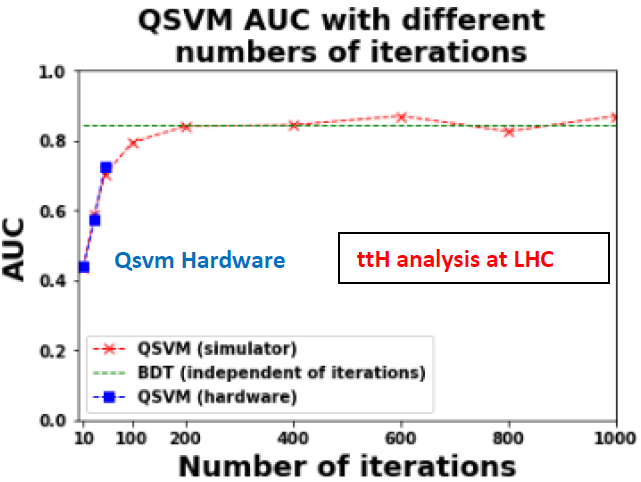}
     \caption{AUC with number of iterations.  Within the limited testing iterations,the  performance  of  the  IBM  Q  quantum  computer  is  compatible  with  the  one  from the quantum simulator~\cite{Chan:2019zwk}. }\label{fig_qsvmibmq}
   \end{minipage}
\end{figure}

The method used in~\cite{Chan:2019zwk} and \cite{terashi2020event} is based on variational quantum algorithms for machine learning (VQML). 
The VQML approach exploits the mapping of input data to an exponentially large quantum state space to enhance the ability to find an optimal solution. 
The data encoding circuit $U_{\Phi(\vec{x})}$ maps the data $\vec{x} \in \Omega$ to the quantum state $\ket{\Phi(\vec{x})}=U_{\Phi(\vec{x})} \ket{0}$.
The quantum state with encoded input data is processed by applying quantum gates to create an ansatz state, which is then measured to produce the output.
The variational quantum circuit $W(\vec{\theta})$ parameterized by $\vec{\theta}$ is applied  ~\cite{Havlcek2019SupervisedLW}
\begin{equation}
W(\vec{\theta}) = U_{\text{loc}}^{(l)}(\theta_l) \;U_{\text{ent}} \ldots U_{\text{loc}}^{(2)}(\theta_2) \; U_{\text{ent}}\; U_{\text{loc}}^{(1)}(\theta_1)
\end{equation}
The probability of outcome $y$ is obtained through
\begin{equation}
%p_y(\vec{x}) \leftarrow \bra{\Phi(\vec{x})} W^\dag(\vec{\theta}) M_y W(\vec{\theta}) \ket{\Phi(\vec{x})}~\cite{Havlcek2019SupervisedLW}
p_y(\vec{x}) \leftarrow \langle \Phi(\vec{x}) | W^\dag(\vec{\theta}) M_y W(\vec{\theta}) | \Phi(\vec{x}) \rangle
\end{equation}
whereas $\{ M_y\}$ is the binary measurement.
%The learning is carried out by tuning the quantum gates such that the output can reproduce the target label of a classification task.
The optimization process consists in learning $\vec{\theta}$ to minimize the loss quantified as a difference between the predicted $p_y(\vec{x})$ and the known classification label $y$. 
Different optimizers, such as COBYLA~\cite{COBYLA} and SPSA~\cite{SPSA1997,SPSA2020}, can be applied.

In~\cite{Chan:2019zwk}, the authors made some promising progress by obtaining preliminary results in the application of IBM quantum simulators and IBM Q quantum computer to ttH (Higgs coupling to top quark pairs) data analysis. 
The authors have measured the AUC (area  under the ROC curve) with different numbers of events in the training dataset.
With  5 qubits and 800 events, the VQML have obtained very close performance to the one obtained using the classical machine learning method BDT (see Figure~\ref{fig_qsvmroc}).
A preliminary test was to perform VQML on the IBM Q quantum computer with 5 qubits, 100 training events and 100 test events. 
Within the limited testing iterations, the performance of the IBM Q quantum computer is compatible with the one from the quantum simulator, which reaches a performance similar to the BDT method with enough iterations (see Figure~\ref{fig_qsvmibmq}).

\begin{figure}[b!]
   \begin{minipage}{0.48\textwidth}
     \centering
     \includegraphics[scale=0.5]{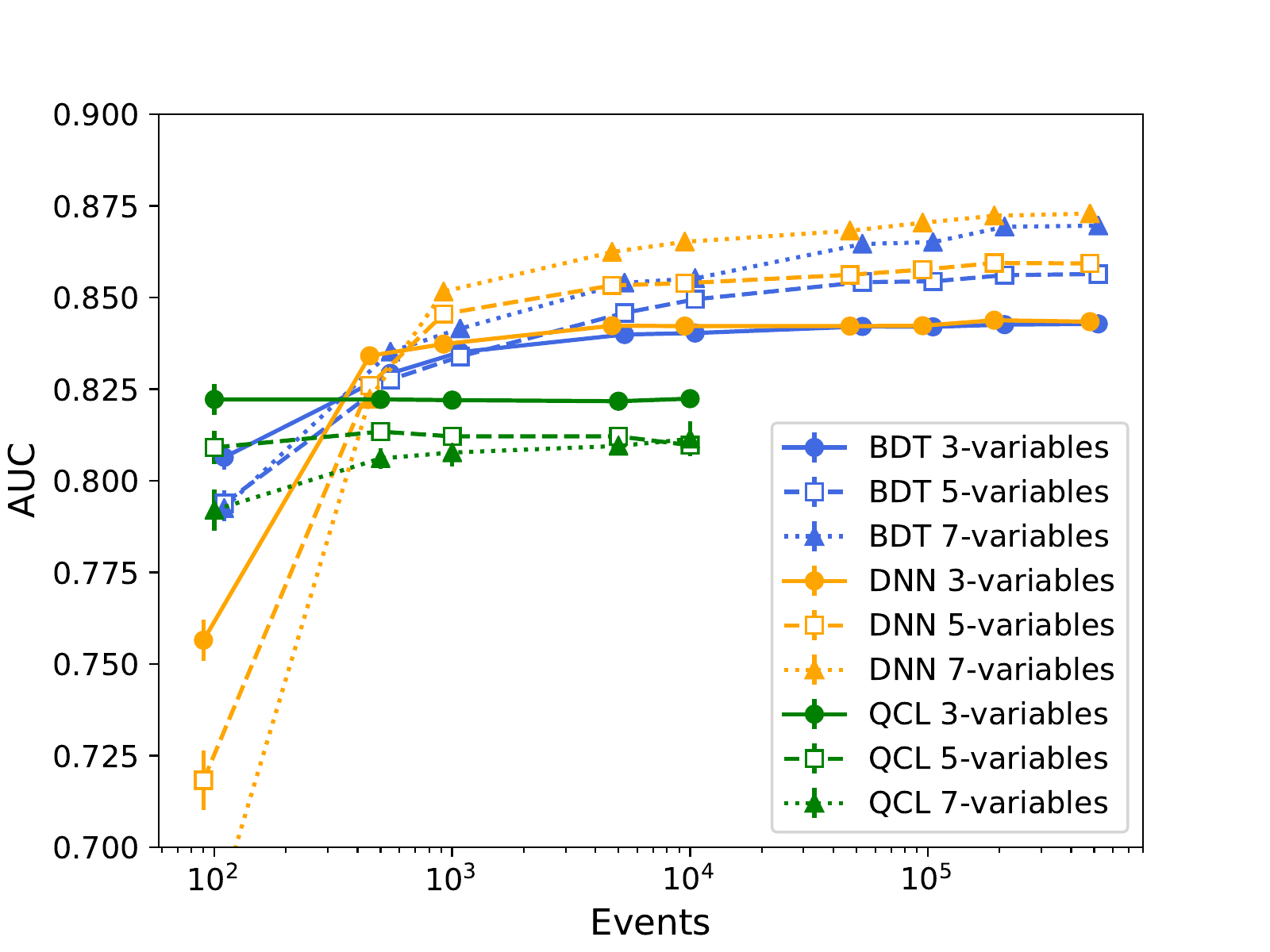}
     \caption{Average AUC values as a function of the training sample size for the BDT, DNN and QCL algorithms with 
3, 5 and 7 variables~\cite{terashi2020event}.}
   \label{fig:VQC1}
   \end{minipage}\hfill
   \begin{minipage}{0.48\textwidth}
     \centering
     \includegraphics[scale=0.5]{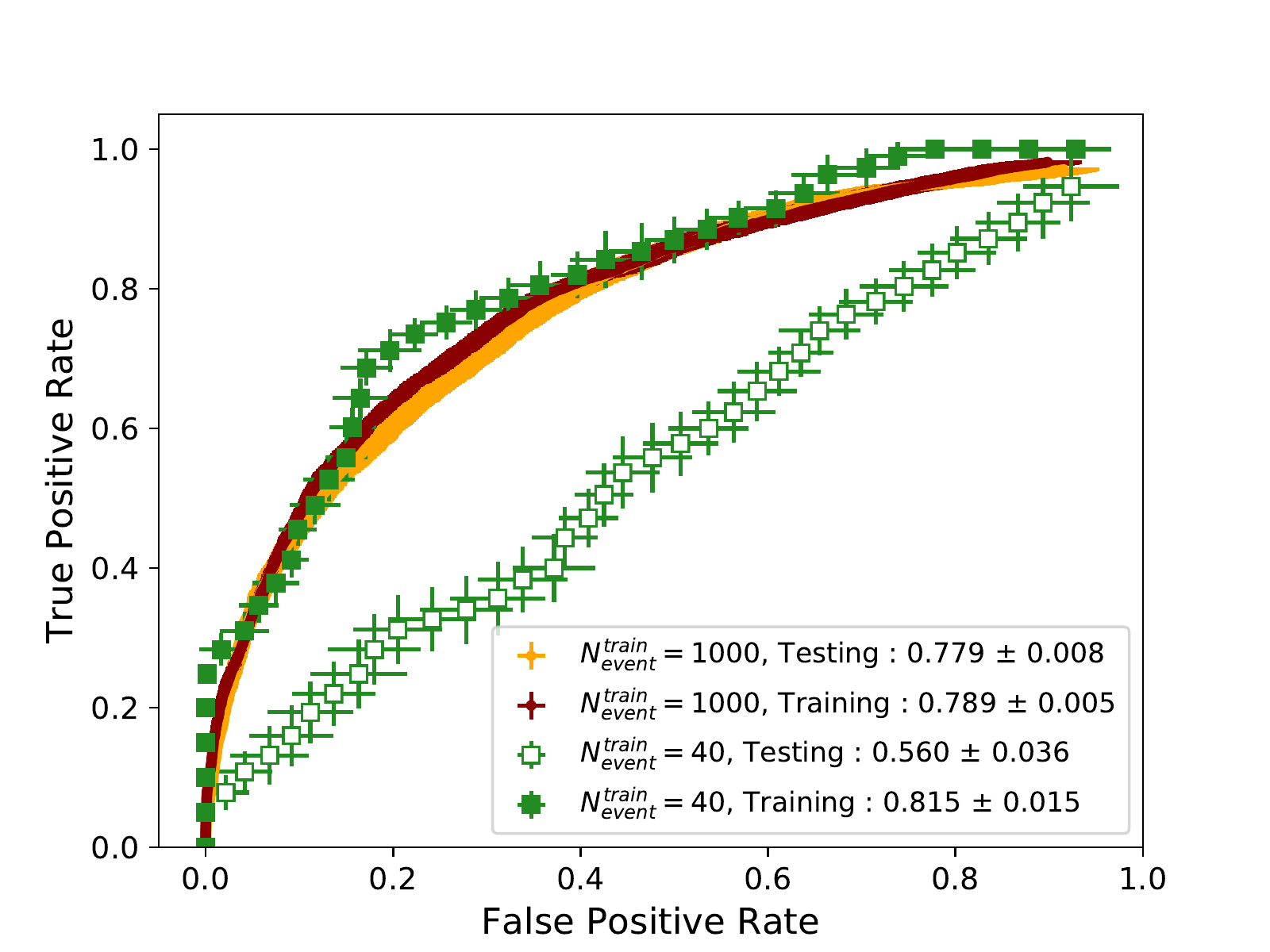}
     \caption{ROC curves in the training and testing phases of the VQC algorithm with 3 variables and 
the training sets of 40 and 1,000 events, obtained using QSM simulator~\cite{terashi2020event}.}
   \label{fig:VQC2}
   \end{minipage}
\end{figure}

In~\cite{terashi2020event}, the authors have attempted to use the VQML algorithm for the classification of a new physics signal predicted in a theory of Supersymmetry. 
Two implementations of the VQML algorithm are tested, the first one called Quantum Circuit Learning (QCL)~\cite{Mitarai_2018}, which is used with the Qulacs simulator~\cite{Qulacs}, and the second called Variational Quantum Classification (VQC)~\cite{Havlcek2019SupervisedLW}, which is used with the QASM simulator and real quantum computing devices. 
The QCL (VQC) uses the combination of $R_Y$ and $R_Z$ (Hadamard and $R_Z$) gates for encoding the input data. 
For the creation of an ansatz state, the combination of an entangling gate and single-qubit rotation gates are used for both implementations. 
The QCL uses the time-evolution gate $e^{-iHt}$ with the Hamiltonian $H$ of an Ising model with random coefficients as an entangling gate while the VQC uses the Hadamard and CNOT gates for that. 
The rotation angles used to create the ansatz are parameters to be tuned, and the number of parameters is chosen to be 27, 45 and 63 for the QCL and 12, 20 and 28 for the VQC using 3, 5 and 7 variables, respectively.

\begin{table}[!htb]
\centering
\caption{AUC values in a training phase for the VQC algorithm running on quantum computers and 
QASM simulator. The training condition is fixed to 3 variables, 40 training events and the number of 
iterations of 100~\cite{terashi2020event}.}
\label{tab:VQC}
\begin{tabular}{ll} \hline
Device/Condition & AUC  \\ \hline
Quantum Computer (Johannesburg) &  $0.799\pm0.020$\\
Quantum Computer (Boeblingen) &  $0.807\pm0.010$\\
QASM simulator & $0.815\pm0.015$\\ \hline
\end{tabular}
\end{table}

The experimental test of the quantum algorithm is performed in \cite{terashi2020event} with the SUSY data set in the UC Irvine Machine Learning Repositiory~\cite{Dua:2019} using cloud Linux servers for the QCL and a local machine and the IBM Q quantum computer for the VQC. 
The performance of the quantum algorithm is compared with BDT and DNN optimized to avoid over-training at each training set. 
%Figure~\ref{fig:VQC1} shows the comparison of the AUC values as a function of the size of the training set for the QCL, BDT and DNN.
The QCL performance is relatively flat in the training size (see Figure~\ref{fig:VQC1})while the performance of  the BDT and DNN improves with the size. 
The computational resource needed to simulate QCL with 10,000 events or more is beyond the capacity used in \cite{terashi2020event}. 
According to these simulation studies, the three algorithms appear to have a comparable discriminating power when restricting the training set 
to be less than $\sim10,000$ events, with an indication that the quantum algorithm might have an advantage
with a small sample of ${\cal O}(100)$ events. 
Figure~\ref{fig:VQC1} shows ROC curves obtained using the 3-variable VQC algorithm on the QASM simulator with different numbers of events in the training set. 
The over-training is clearly visible if the training set contains only 40 events while it is largely gone when the training set is increased to 1,000. 
The small sample of 40 events is used to train the VQC model with IBM Q quantum computers as well. The AUC values from the QASM simulator and quantum computers are given in table~\ref{tab:VQC}. 
The results from the quantum computers appear to be slightly worse than those from the simulator, though they are consistent within the uncertainties (defined as the standard deviations of five measurements). The authors of~\cite{terashi2020event} conclude that 
the variational quantum circuit can learn the properties of the input data with real quantum device, acquiring classification power for physics events of interest.

\section{Applications Coming Soon}\label{sec:comingsoon}

An interesting line of research concerns generative models, such as Boltzmann Machines, Variational Auto-Encoders and Generative Adversarial Networks, and their quantum counterparts. 
Classical generative models are being investigated by the HEP community as solutions to speed up Monte-Carlo simulation, because of their ability to model complex probability distributions, and the relative lower computation cost during the prediction phase. 
Training those models is, however, a difficult task, and computing intensive.
Coverage is one of the major issues when training or validating generative models performance and it is related to their representational power and how it maps to the original probability distribution. 
From this point of view quantum generative models might show an advantage, while relieving the computational cost \cite{Gaoeaat9004}. 

Quantum SVMs (Support Vector Machines) offer an attractive approach not fully exploited in HEP.
An SVM ~\cite{classicalsvm} is a supervised machine learning method which outputs an optimal hyperplane to categorize samples between two classes to classify data points. 
A quantum-enhanced kernel for SVM ~\cite{Havlcek2019SupervisedLW} can map the input vectors to an exponential Hilbert space, which could make it easier to construct an optimal hyperplane and increase the classification performance. 
Additionally, to calculate the quantum-enhanced kernel, the number of circuits is a function of the square of the number of input vectors, which may not be a good selection for classifying huge number of events.
Multiple groups are actively exploring quantum kernel methods with gate-based quantum computers for event classification.
Currently, these methods are limited by the dimensionality reduction required to make data compatible with modern hardware.
Studying these algorithms provides new and different insights into the performance of modern computing platforms though.
For example, they compute data element overlaps in Hilbert space, and the outcome state distributions are sensitive to device noise in different ways than variational algorithms like VQE or QAOA.
New schemes for approaching \textit{quantum feature map} in particular ~\cite{lloyd2020quantum} are interesting directions.

\section{Discussion and Outlook}\label{sec:discussion}

When considering applications of quantum machine learning for a field such as high energy physics (HEP), the immediate question is whether we have reason to believe that quantum machine learning -- for near-term \textit{or} universal quantum computers -- is particularly suited to this type of application. 
The truth is that it is simply too early to tell, and only further investigation of the methods will provide the answers.

One feature of HEP data sets is that they are notoriously large. 
In principle, this makes quantum speed-ups attractive, as they could be crucial to analyse big amounts of data.
But significant (that is, exponential) speed-ups in quantum machine learning are still controversial as to their scope \cite{aaronson15} and in some cases, their true quantum nature \cite{tang2019quantum}.
They often rely on special properties of the data such as sparsity \cite{arrazola2019quantum}, or a special oracle or device that can load the data in superposition \cite{kerenedis16}.
The appeal of near-term approaches to quantum machine learning is without doubt that ideas can be easily tested on a small scale, using the rich landscape of quantum programming languages, cloud-based quantum computers, and quantum machine learning software packages. 
Even so, to encode large data sets into a quantum system to sufficient precision and to measure the outputs for every events in the dataset is a physical challenge that is significantly out of the scope of near-term quantum computing. 
Of course, in the age of Big Data, the large size of the data sets are not unique to HEP, and it needs to be further established whether the intersection discussed in this review poses any \textit{particular} challenge to machine learning which would motivate the use of quantum computers.

\subsection{Experimenting with Quantum Annealers}
%% digitial annelear
Despite continuous improvement of quantum annealers they remain noisy, with limited number of qubits, and limited connectivity.

\subsubsection*{Solver Heuristics.}
A huge challenge is to map the reformulated problem to an actual device with a limited connectivity \cite{benedetti16a}, and it is often necessary to include connectivity constraints already into the loss itself. 
One alternative available in the D-Wave software stack is \textit{qbsolve} \cite{booth2017partitioning}, a heuristic that allows to split large problems in several smaller ones that in turn can be solve on the available hardware.
This allows one to experiment with much larger QUBO than the one directly solvable with existing hardware, but in return requires additional computing resources. 
It also prevents us from directly probing the stand-alone capabilities of the device.

\subsubsection*{Digital Devices.}
Digital annealers \cite{digitalannealer} offer the potential to prototype algorithms with large numbers of \textit{digital qubits}.
Using custom ASICs, digital annealers are capable of simulating fully-connected quantum annealers with 4,096 qubits (with 64 bit precision) or as many as 8,192 qubits (with 16 bit precision).
In principle, a digital annealer cluster could offer up to 1,000,000 qubits using multi-chip support.
While in the very long-run fully quantum annealers should be able to overtake digital simulators, in the near-term, these machines are exciting application test-beds and may even be able to deliver competitive results.

\subsection{Experimenting with Quantum Circuits}

Applying quantum algorithms on quantum hardware is the core aim at any research on quantum computing.
But the scale of even state-of-the-art studies quickly reveals the limitations of current-day hardware. 
Typical implementations use only a few qubits and datasets of four features (for example, \cite{havlivcek2019supervised, benedetti2019generative, schuld2017implementing}). 
The limited number of qubits, connectivity and short decoherence time of the current quantum hardware make it difficult to experiment with large and long variational circuits.

\subsubsection*{Circuit Architecture.}
% circuit optimization
In the papers reviewed above, the quantum circuit architecture (the types and numbers of gates) is fixed and only the parameters of the gates are optimized.
In combination of this approach, search for optimal gate assembly is also possible.
In \cite{mckiernan2019automated}, reinforcement learning is used to derive circuits to solve combinatorial problems.
This technique might provide further handle at developing well performing quantum machine learning models.

\subsubsection*{Error Mitigation.}
% error mitigation
Practically, circuit-based qubit devices allow only a few gates to be performed before a signal is drowned in noise. 
%Error mitigation \cite{temme2017error} strategies, which aim at improving noisy results in the absence of actual error correction during the quantum computation, require increased resources.
The fidelity of measurements on quantum device can be improved via error mitigation strategies~\cite{temme2017error}. 
Various techniques allow experiments with an increased number of gates or better qubit connectivity.
In addition to techniques making explicit assumption on the form and origin of the noise, machine learning approaches can be used to learn directly from the device-dependent noise.
The integration of such noise-modelling-cancelling technique of circuit compiler would help with experimenting on quantum device, at the cost of increased resources.

%% simulator time
\subsubsection*{Circuit Simulation.}
Prototyping quantum algorithms with a classical simulator is an important step in the development and testing of new algorithms. 
The classical simulator used for the VQML study in~\cite{terashi2020event} has enabled the authors to test the QCL algorithm with up to seven variables or $\sim10,000$ events for the training set size. 
The simulation time and memory usage increases exponentially with the number of input variables in the creation of variational quantum states with $W(\vec{\theta})$. %, which becomes a bottleneck in the simulation of the VQML algorithm. 
Despite continuous improvement in the simulator \cite{Qulacs}, the experimentation with circuits with large number of qubits is still hampered by this computation requirement.
Of course, it is expected that the simulation of a quantum device will be classically hard. 
%By nature, the quantum device would solve this by evolving under the laws of quantum mechanics.
Because of this, it may be better when possible to experiment on smaller numbers of qubits ---where circuits can be run--- and study the time to solution or complexity, as a function of the number of qubits.

\subsubsection*{Optimization in Quantum Machine Learning.}

There are two types of optimizer: gradient-based and derivative-free. 
For some derivative-free optimizers, it may require many iterations to achieve good training performance as the number of variational parameters increases.  
For the gradient-based optimizer, fewer iterations may be required. However, to calculate the gradient is also difficult~\cite{analyticgradient} and numerical differentiation requires the circuit to be run additional times as the number of variational parameters increases.
Changing a single circuit parameter for the evaluation of gradients through a cloud-based service can take of the order of many seconds, which quickly makes optimization of even a small system a matter of hours and days. 

%\subsubsection*{Data Ingestion.}
%The required time to encode an attribute into a quantum system and the output time to measure a quantum state, including waves preparing time and waves readout time,   is still much larger than the processing time spend on the quantum process unit (QPU). 
%For HEP classification that requires a lot of encoding and measuring, further technological improvements, such as quantum memory, may be necessary to cope with large datasets.

\subsection{Quantum data}

All the algorithms described in this review so far made use of a classical machine learning dataset, embedded into a quantum device. However, quantum machine learning algorithms have the unique property to be usable with a dataset made of quantum states, or \textit{quantum data} \cite{unsupervised-quantum-data, no-free-lunch}. Those input quantum states are usually the output of some quantum circuits (e.g. circuits that extract the ground state of different Hamiltonians \cite{qcnn}) and are then processed by a variational circuit that has learned a desired quantum function (e.g. a property of this ground-state). However, one could also imagine directly feeding the quantum objects resulting from a HEP, dark matter, or gravitational wave detection experiment into the QML algorithm.
Several application of machine learning on quantum data have been developed, including clustering of quantum states \cite{unsupervised-quantum-data}, detecting anomalies on a quantum device \cite{quantum-anomaly-detection}, learning algorithms to estimate the fidelity or the purity of a state \cite{learning-purity, quantum-fidelity-estimation}, learning phases of matter \cite{qcnn} and classifying quantum states \cite{classification-quantum-data, learning-coherent-states}.

The question of how to exploit the quantum nature of the systems generating HEP data has not been prominent in the literature, but there are two interesting outlooks. 

The first is to do quantum machine learning \textit{directly} on the quantum objects measured in HEP. 
As an example, instead of processing the classical signal formed in photonic sensors, one could direct the photons into a \textit{photonic quantum computer} and apply a variational circuit before conducting the final measurement. 
The circuit could be trained to extract important information from the quantum state, or to classify the state. 
Applying this process to axion dark matter experiments \cite{doi:10.1063/PT.3.4227} or to neutrino detectors \cite{linear-response} could be promising research directions.

The second path follows the idea of \textit{quantum simulations} \cite{lloyd1996universal,RevModPhys.86.153}, an important use of quantum computers in simulating complex quantum systems to determine their properties. 
If a HEP experiment could be simulated on a quantum computer \cite{preskill-qft-simulation,linear-response,neutrino-nucleus}, the simulation could be followed by a quantum machine learning routine executed on the very same device, and analysing the quantum states produced by the simulation. 
Instead of costly state tomography to characterise the results, the wave function is directly accessed and important information extracted.

In both cases, an important insight from quantum machine learning --- possibly the one with the highest future impact on other quantum disciplines --- is the ability to differentiate through quantum computations. 
This includes a wealth of knowledge and practical methods to get partial derivatives of a measurement result with respect to (classical) physical parameters of the experiment, such as a magnetic field strength or pulse length.
Quantum differentiation opens a door to design experiments by adaptively optimizing some cost functions, which is crucial for quantum data analysis.

\subsection{Concluding Remarks}

Overall, we are just at the beginning of exploring the intersection between quantum machine learning and high energy physics. 
The papers presented in this review therefore have to be understood as exploratory studies that propose angles to approach the problem of how to use quantum machine learning algorithms to understand fundamental particles. 

We presented papers on performing classification using quantum machine learning with quantum annealing, restrictive Boltzmann machines, quantum graph networks and variational quantum circuits.
The capacity of quantum annealers to perform classification is limited due to the restrictive formulation of the problem.
Quantum-circuit-based machine learning is yet of limited performance due to the necessary down-scaling of the problems, so as to fit on the quantum device, or to be amenable in simulation. 

In the outlook we discussed practical considerations of experimenting with quantum machine learning and the prospect of analysing quantum data. 
These challenges put quantum machine learning into a particularly difficult spot. 
The quality of a machine learning algorithm is usually estimated through empirical benchmarks on pseudo-realistic datasets. 
Evidence from deep learning suggests that machine learning on large datasets behaves very differently from the small-data regime. 
And while consistently improving, the theory of machine learning is currently unable to explain the performance of algorithms such as neural networks. 
The challenges for practical experiments as well as fundamental limits of classical simulations restrict quantum machine learning benchmarks to small proof-of-principle investigations that may only say very little about their performance in realistic settings.

As the technology develops, more theory work is needed to understand the power of near-term quantum machine learning.
While the current performance of quantum machine learning on high energy physics data is limited, there is hope that future advances on both quantum devices and quantum algorithms will help with the computation challenges of particle physics.

\section*{Acknowledgement}
The authors wish to thank Alexander Zlokapa, Joshua Job, Cenk Tuysuz and Shaojun Sun for sharing material reproduced here.

JRV is partially supported by DOE/HEP QuantISED program grant, Quantum Machine Learning and Quantum Computation Frameworks (QMLQCF) for HEP, award number DE-SC0019227. 
MS acknowledges support by the Big Data and Informatics Flagship and BDSS initiative of the University of KwaZulu-Natal.
WG is partially supported by DOE/HEP QuantISED program grant, Application of Quantum Machine Learning to High Energy Physics Analysis at the LHC using IBM Quantum Computer Simulator and Hardware, award number DE-SC0020416.
GP is partially supported by the DOE/HEP QuantISED program grant HEP Machine Learning and Optimization Go Quantum, identification number 0000240323.
This manuscript has been authored by Fermi Research Alliance, LLC under Contract No. DE-AC02-07CH11359 with the U.S. Department of Energy, Office of Science, Office of High Energy Physics.

Data sharing is not applicable to this article as no new data were created or analysed in this study.

\vspace{1cm}

\bibliography{QMLinHEP}

\begin{thebibliography}{100}

\bibitem{Radovic:2018dip}
Alexander Radovic, Mike Williams, David Rousseau, Michael Kagan, Daniele
  Bonacorsi, Alexander Himmel, Adam Aurisano, Kazuhiro Terao, and Taritree
  Wongjirad.
\newblock {Machine learning at the energy and intensity frontiers of particle
  physics}.
\newblock {\em Nature}, 560(7716):41--48, 2018.

\bibitem{Albertsson:2018maf}
Kim Albertsson et~al.
\newblock {Machine Learning in High Energy Physics Community White Paper}.
\newblock {\em J. Phys. Conf. Ser.}, 1085(2):022008, 2018.

\bibitem{Guest:2018yhq}
Dan Guest, Kyle Cranmer, and Daniel Whiteson.
\newblock {Deep Learning and its Application to LHC Physics}.
\newblock {\em Ann. Rev. Nucl. Part. Sci.}, 68:161--181, 2018.

\bibitem{Shapoval:2019txi}
Illya Shapoval and Paolo Calafiura.
\newblock {Quantum Associative Memory in HEP Track Pattern Recognition}.
\newblock {\em EPJ Web Conf.}, 214:01012, 2019.

\bibitem{Bapst:2019llh}
Frederic Bapst, Wahid Bhimji, Paolo Calafiura, Heather Gray, Wim Lavrijsen, and
  Lucy Linder.
\newblock {A pattern recognition algorithm for quantum annealers}.
\newblock 2 2019.

\bibitem{Bauer:2019qxa}
Christian~W. Bauer, Wibe~A. De~Jong, Benjamin Nachman, and Davide Provasoli.
\newblock {A quantum algorithm for high energy physics simulations}.
\newblock 4 2019.

\bibitem{Zlokapa:2019tkn}
Alexander Zlokapa, Abhishek Anand, Jean-Roch Vlimant, Javier~M. Duarte, Joshua
  Job, Daniel Lidar, and Maria Spiropulu.
\newblock {Charged Particle Tracking with Quantum Annealing-Inspired
  Optimization}.
\newblock 8 2019.

\bibitem{Cormier:2019kcq}
Kyle Cormier, Riccardo Di~Sipio, and Peter Wittek.
\newblock {Unfolding measurement distributions via quantum annealing}.
\newblock {\em JHEP}, 11:128, 2019.

\bibitem{feynman}
Richard~P. Feynman.
\newblock Simulating physics with computers.
\newblock {\em International Journal of Theoretical Physics}, 21(6):467--488,
  1982.

\bibitem{qaoa}
Edward Farhi, Jeffrey Goldstone, and Sam Gutmann.
\newblock A quantum approximate optimization algorithm, 2014.

\bibitem{brandao-sdp}
F.~G. S.~L. {Brandao} and K.~M. {Svore}.
\newblock Quantum speed-ups for solving semidefinite programs.
\newblock In {\em 2017 IEEE 58th Annual Symposium on Foundations of Computer
  Science (FOCS)}, pages 415--426, 2017.

\bibitem{childs-convex-optimization}
Shouvanik Chakrabarti, Andrew~M. Childs, Tongyang Li, and Xiaodi Wu.
\newblock Quantum algorithms and lower bounds for convex optimization.
\newblock {\em {Quantum}}, 4:221, January 2020.

\bibitem{quantum-chemistry-1}
Yudong Cao, Jonathan Romero, Jonathan~P. Olson, Matthias Degroote, Peter~D.
  Johnson, M{\'a}ria Kieferov{\'a}, Ian~D. Kivlichan, Tim Menke, Borja
  Peropadre, Nicolas P.~D. Sawaya, Sukin Sim, Libor Veis, and Al{\'a}n
  Aspuru-Guzik.
\newblock Quantum chemistry in the age of quantum computing.
\newblock {\em Chemical Reviews}, 119(19):10856--10915, 2019.

\bibitem{quantum-chemistry-2}
Sam McArdle, Suguru Endo, Alan Aspuru-Guzik, Simon Benjamin, and Xiao Yuan.
\newblock Quantum computational chemistry, 2018.

\bibitem{qml-1}
Jacob Biamonte, Peter Wittek, Nicola Pancotti, Patrick Rebentrost, Nathan
  Wiebe, and Seth Lloyd.
\newblock \href{https://doi.org/10.1038/nature23474}{Quantum machine learning}.
\newblock {\em Nature}, 549:195 EP, 2017.

\bibitem{qml-2}
Peter Wittek.
\newblock {\em Quantum Machine Learning: What Quantum Computing Means to Data
  Mining}.
\newblock Academic Press, 2014.

\bibitem{qml-3}
Maria Schuld and Francesco Petruccione.
\newblock {\em Supervised Learning with Quantum Computers}.
\newblock Springer International Publishing, 2018.

\bibitem{linear-response}
Alessandro Roggero and Joseph Carlson.
\newblock Dynamic linear response quantum algorithm.
\newblock {\em Phys. Rev. C}, 100:034610, Sep 2019.

\bibitem{neutrino-nucleus}
Alessandro Roggero, Andy C.~Y. Li, Joseph Carlson, Rajan Gupta, and Gabriel~N.
  Perdue.
\newblock Quantum computing for neutrino-nucleus scattering.
\newblock {\em Phys. Rev. D}, 101:074038, Apr 2020.

\bibitem{qc-atomic-nucleus}
Eugene~F Dumitrescu, Alex~J McCaskey, Gaute Hagen, Gustav~R Jansen, Titus~D
  Morris, T~Papenbrock, Raphael~C Pooser, David~Jarvis Dean, and Pavel
  Lougovski.
\newblock Cloud quantum computing of an atomic nucleus.
\newblock {\em Physical review letters}, 120(21):210501, 2018.

\bibitem{preskill-qft-simulation}
John Preskill.
\newblock Simulating quantum field theory with a quantum computer.
\newblock {\em arXiv preprint arXiv:1811.10085}, 2018.

\bibitem{lattice-gauge-theory}
Esteban~A Martinez, Christine~A Muschik, Philipp Schindler, Daniel Nigg,
  Alexander Erhard, Markus Heyl, Philipp Hauke, Marcello Dalmonte, Thomas Monz,
  Peter Zoller, et~al.
\newblock Real-time dynamics of lattice gauge theories with a few-qubit quantum
  computer.
\newblock {\em Nature}, 534(7608):516--519, 2016.

\bibitem{subatomic-many-body}
Hsuan-Hao Lu, Natalie Klco, Joseph~M Lukens, Titus~D Morris, Aaina Bansal,
  Andreas Ekstr{\"o}m, Gaute Hagen, Thomas Papenbrock, Andrew~M Weiner,
  Martin~J Savage, et~al.
\newblock Simulations of subatomic many-body physics on a quantum frequency
  processor.
\newblock {\em Physical Review A}, 100(1):012320, 2019.

\bibitem{non-abelian-gauge-field}
Natalie Klco, Martin~J Savage, and Jesse~R Stryker.
\newblock Su (2) non-abelian gauge field theory in one dimension on digital
  quantum computers.
\newblock {\em Physical Review D}, 101(7):074512, 2020.

\bibitem{deutsch}
David Deutsch.
\newblock Quantum theory, the church{\textendash}turing principle and the
  universal quantum computer.
\newblock {\em Proceedings of the Royal Society of London. A. Mathematical and
  Physical Sciences}, 400(1818):97--117, 1985.

\bibitem{shor-algorithm}
Peter~W. Shor.
\newblock Polynomial-time algorithms for prime factorization and discrete
  logarithms on a quantum computer.
\newblock {\em {SIAM} Journal on Computing}, 26(5):1484--1509, 1997.

\bibitem{grover-algorithm}
Lov~K. Grover.
\newblock A fast quantum mechanical algorithm for database search.
\newblock In {\em Proceedings of the Twenty-Eighth Annual ACM Symposium on
  Theory of Computing}, STOC ’96, page 212–219, New York, NY, USA, 1996.
  Association for Computing Machinery.

\bibitem{hhl-algorithm}
Aram~W. Harrow, Avinatan Hassidim, and Seth Lloyd.
\newblock Quantum algorithm for linear systems of equations.
\newblock {\em Phys. Rev. Lett.}, 103:150502, Oct 2009.

\bibitem{corcoles2019challenges}
Antonio~D C{\'o}rcoles, Abhinav Kandala, Ali Javadi-Abhari, Douglas~T McClure,
  Andrew~W Cross, Kristan Temme, Paul~D Nation, Matthias Steffen, and
  JM~Gambetta.
\newblock Challenges and opportunities of near-term quantum computing systems.
\newblock {\em arXiv preprint arXiv:1910.02894}, 2019.

\bibitem{gosset-quantum-mean-value}
Sergey Bravyi, David Gosset, and Ramis Movassagh.
\newblock Classical algorithms for quantum mean values, 2019.

\bibitem{Preskill_2018}
John Preskill.
\newblock Quantum computing in the nisq era and beyond.
\newblock {\em Quantum}, 2:79, Aug 2018.

\bibitem{schuld18cc}
Maria Schuld, Alex Bocharov, Krysta Svore, and Nathan Wiebe.
\newblock Circuit-centric quantum classifiers.
\newblock {\em arXiv preprint arXiv:1804.00633}, 2018.

\bibitem{nishimori-qa}
Tadashi Kadowaki and Hidetoshi Nishimori.
\newblock Quantum annealing in the transverse ising model.
\newblock {\em Phys. Rev. E}, 58:5355--5363, Nov 1998.

\bibitem{mit-qa}
Edward Farhi, Jeffrey Goldstone, Sam Gutmann, and Michael Sipser.
\newblock Quantum computation by adiabatic evolution, 2000.

\bibitem{adiabatic-grover}
J\'er\'emie Roland and Nicolas~J. Cerf.
\newblock Quantum search by local adiabatic evolution.
\newblock {\em Phys. Rev. A}, 65:042308, Mar 2002.

\bibitem{biamonte17}
Jacob Biamonte, Peter Wittek, Nicola Pancotti, Patrick Rebentrost, Nathan
  Wiebe, and Seth Lloyd.
\newblock Quantum machine learning.
\newblock {\em Nature}, 549(7671):195, 2017.

\bibitem{arunachalam2017guest}
Srinivasan Arunachalam and Ronald de~Wolf.
\newblock Guest column: A survey of quantum learning theory.
\newblock {\em ACM SIGACT News}, 48(2):41--67, 2017.

\bibitem{ciliberto2020fast}
Carlo Ciliberto, Andrea Rocchetto, Alessandro Rudi, and Leonard Wossnig.
\newblock Fast quantum learning with statistical guarantees.
\newblock {\em arXiv preprint arXiv:2001.10477}, 2020.

\bibitem{lloyd14}
Seth Lloyd, Masoud Mohseni, and Patrick Rebentrost.
\newblock Quantum principal component analysis.
\newblock {\em Nature Physics}, 10:631--633, 2014.

\bibitem{rebentrost2014quantum}
Patrick Rebentrost, Masoud Mohseni, and Seth Lloyd.
\newblock Quantum support vector machine for big data classification.
\newblock {\em Physical Review Letters}, 113(13):130503, 2014.

\bibitem{kerenedis16}
Iordanis Kerenedis and Anupam Prakash.
\newblock Quantum recommendation systems.
\newblock In {\em Kerenidis, Iordanis, and Anupam Prakash. "Quantum
  Recommendation Systems." LIPIcs-Leibniz International Proceedings in
  Informatics.}, volume~67, 2017.

\bibitem{ciliberto2018quantum}
Carlo Ciliberto, Mark Herbster, Alessandro~Davide Ialongo, Massimiliano Pontil,
  Andrea Rocchetto, Simone Severini, and Leonard Wossnig.
\newblock Quantum machine learning: a classical perspective.
\newblock {\em Proceedings of the Royal Society A: Mathematical, Physical and
  Engineering Sciences}, 474(2209):20170551, 2018.

\bibitem{perdomo2018opportunities}
Alejandro Perdomo-Ortiz, Marcello Benedetti, John Realpe-G{\'o}mez, and Rupak
  Biswas.
\newblock Opportunities and challenges for quantum-assisted machine learning in
  near-term quantum computers.
\newblock {\em Quantum Science and Technology}, 3(3):030502, 2018.

\bibitem{farhi18}
Edward Farhi and Hartmut Neven.
\newblock Classification with quantum neural networks on near term processors.
\newblock {\em arXiv preprint arXiv:1802.06002}, 2018.

\bibitem{benedetti16b}
Marcello Benedetti, John Realpe-G{\'o}mez, Rupak Biswas, and Alejandro
  Perdomo-Ortiz.
\newblock Quantum-assisted learning of hardware-embedded probabilistic
  graphical models.
\newblock {\em Physical {R}eview {X}}, 7:041052, 2017.

\bibitem{neven08bin}
Hartmut Neven, Vasil~S Denchev, Geordie Rose, and William~G Macready.
\newblock Training a binary classifier with the quantum adiabatic algorithm.
\newblock {\em arXiv preprint arXiv:0811.0416}, 2008.

\bibitem{lloyd2020quantum}
Seth Lloyd, Maria Schuld, Aroosa Ijaz, Josh Izaac, and Nathan Killoran.
\newblock Quantum embeddings for machine learning.
\newblock {\em arXiv preprint arXiv:2001.03622}, 2020.

\bibitem{mcclean16}
Jarrod~R McClean, Jonathan Romero, Ryan Babbush, and Al{\'a}n Aspuru-Guzik.
\newblock The theory of variational hybrid quantum-classical algorithms.
\newblock {\em New Journal of Physics}, 18(2):023023, 2016.

\bibitem{mitarai2018quantum}
Kosuke Mitarai, Makoto Negoro, Masahiro Kitagawa, and Keisuke Fujii.
\newblock Quantum circuit learning.
\newblock {\em Physical Review A}, 98(3):032309, 2018.

\bibitem{schuld2019evaluating}
Maria Schuld, Ville Bergholm, Christian Gogolin, Josh Izaac, and Nathan
  Killoran.
\newblock Evaluating analytic gradients on quantum hardware.
\newblock {\em Physical Review A}, 99(3):032331, 2019.

\bibitem{bergholm2018pennylane}
Ville Bergholm, Josh Izaac, Maria Schuld, Christian Gogolin, and Nathan
  Killoran.
\newblock Pennylane: Automatic differentiation of hybrid quantum-classical
  computations.
\newblock {\em arXiv preprint arXiv:1811.04968}, 2018.

\bibitem{broughton2020tensorflow}
Michael Broughton, Guillaume Verdon, Trevor McCourt, Antonio~J Martinez,
  Jae~Hyeon Yoo, Sergei~V Isakov, Philip Massey, Murphy~Yuezhen Niu, Ramin
  Halavati, Evan Peters, et~al.
\newblock Tensorflow quantum: A software framework for quantum machine
  learning.
\newblock {\em arXiv preprint arXiv:2003.02989}, 2020.

\bibitem{harrow2019low}
Aram Harrow and John Napp.
\newblock Low-depth gradient measurements can improve convergence in
  variational hybrid quantum-classical algorithms.
\newblock {\em arXiv preprint arXiv:1901.05374}, 2019.

\bibitem{liu2018differentiable}
Jin-Guo Liu and Lei Wang.
\newblock Differentiable learning of quantum circuit born machines.
\newblock {\em Physical Review A}, 98(6):062324, 2018.

\bibitem{benedetti2019generative}
Marcello Benedetti, Delfina Garcia-Pintos, Oscar Perdomo, Vicente
  Leyton-Ortega, Yunseong Nam, and Alejandro Perdomo-Ortiz.
\newblock A generative modeling approach for benchmarking and training shallow
  quantum circuits.
\newblock {\em npj Quantum Information}, 5(1):1--9, 2019.

\bibitem{lloyd2018quantum}
Seth Lloyd and Christian Weedbrook.
\newblock Quantum generative adversarial learning.
\newblock {\em Physical review letters}, 121(4):040502, 2018.

\bibitem{dallaire2018quantum}
Pierre-Luc Dallaire-Demers and Nathan Killoran.
\newblock Quantum generative adversarial networks.
\newblock {\em Physical Review A}, 98(1):012324, 2018.

\bibitem{neven2009nips}
Harmut Neven, Vasil~S Denchev, Marshall Drew-Brook, Jiayong Zhang, William~G
  Macready, and Geordie Rose.
\newblock Nips 2009 demonstration: Binary classification using hardware
  implementation of quantum annealing.
\newblock {\em Quantum}, pages 1--17, 2009.

\bibitem{Pudenz:2011tue}
Kristen~L. Pudenz and Daniel~A. Lidar.
\newblock {Quantum adiabatic machine learning}.
\newblock {\em Quant. Inf. Proc.}, 12(5):2027--2070, 2013.

\bibitem{glasser2018neural}
Ivan Glasser, Nicola Pancotti, Moritz August, Ivan~D Rodriguez, and J~Ignacio
  Cirac.
\newblock Neural-network quantum states, string-bond states, and chiral
  topological states.
\newblock {\em Physical Review X}, 8(1):011006, 2018.

\bibitem{amin2018quantum}
Mohammad~H Amin, Evgeny Andriyash, Jason Rolfe, Bohdan Kulchytskyy, and Roger
  Melko.
\newblock Quantum boltzmann machine.
\newblock {\em Physical Review X}, 8(2):021050, 2018.

\bibitem{QAML}
Alex Mott, Joshua Job, Jean-Roch Vlimant, Daniel Lidar, and Maria Spiropulu.
\newblock Solving a higgs optimization problem with quantum annealing for
  machine learning.
\newblock {\em Nature}, 550:375--379, 10 2017.

\bibitem{alex2019quantum}
Alexander Zlokapa, Alex Mott, Joshua Job, Jean-Roch Vlimant, Daniel Lidar, and
  Maria Spiropulu.
\newblock Quantum adiabatic machine learning with zooming, 2019.

\bibitem{Caldeira:2019lzf}
João Caldeira, Joshua Job, Steven~H. Adachi, Brian Nord, and Gabriel~N.
  Perdue.
\newblock {Restricted Boltzmann Machines for galaxy morphology classification
  with a quantum annealer}.
\newblock 11 2019.

\bibitem{201230}
S.~Chatrchyan et~al.
\newblock Observation of a new boson at a mass of 125 gev with the cms
  experiment at the lhc.
\newblock {\em Physics Letters B}, 716(1):30 -- 61, 2012.

\bibitem{20121}
G.~Aad et~al.
\newblock Observation of a new particle in the search for the standard model
  higgs boson with the atlas detector at the lhc.
\newblock {\em Physics Letters B}, 716(1):1 -- 29, 2012.

\bibitem{Kirkpatrick671}
S.~Kirkpatrick, C.~D. Gelatt, and M.~P. Vecchi.
\newblock Optimization by simulated annealing.
\newblock {\em Science}, 220(4598):671--680, 1983.

\bibitem{Katzgraber_2006}
Helmut~G Katzgraber, Simon Trebst, David~A Huse, and Matthias Troyer.
\newblock Feedback-optimized parallel tempering monte carlo.
\newblock {\em Journal of Statistical Mechanics: Theory and Experiment},
  2006(03):P03018--P03018, mar 2006.

\bibitem{Hinton2002}
Geoffrey~E. Hinton.
\newblock Training products of experts by minimizing contrastive divergence.
\newblock {\em Neural Computation}, 14(8):1771--1800, 2002.

\bibitem{adachi2015application}
Steven~H. Adachi and Maxwell~P. Henderson.
\newblock Application of quantum annealing to training of deep neural networks,
  2015.

\bibitem{Adachi2015}
Steven~H. {Adachi} and Maxwell~P. {Henderson}.
\newblock Application of quantum annealing to training of deep neural networks.
\newblock arXiv e-print, 2015.

\bibitem{CHEP2019.Cenk}
B.~Dermikoz D. Dobos F. Fracas K. Novotny K. Potamianos S. Vallecorsa
  J.~Vlimant C.~Tuysuz, F.~Carminati.
\newblock Particle track reconstruction with quantum algorithms, 2020.

\bibitem{Chan:2019zwk}
Jay Chan, Wen Guan, Shaojun Sun, Alex~Zeng Wang, Sau~Lan Wu, Chen Zhou, Miron
  Livny, Federico Carminati, and Alberto~Di Meglio.
\newblock Application of quantum machine learning to high energy physics
  analysis at lhc using ibm quantum computer simulators and ibm quantum
  computer hardware.
\newblock {\em PoS}, LeptonPhoton2019:049, 2019.

\bibitem{terashi2020event}
Koji Terashi, Michiru Kaneda, Tomoe Kishimoto, Masahiko Saito, Ryu Sawada, and
  Junichi Tanaka.
\newblock Event classification with quantum machine learning in high-energy
  physics, 2020.

\bibitem{hl-lhc}
T.~Nakamoto L.~Rossi G.~Apollinari, O.~Bruening.
\newblock High luminosity large hadron collider hl-lhc, 2017.

\bibitem{TTN}
Edward Grant, Marcello Benedetti, Shuxiang Cao, Andrew Hallam, Joshua Lockhart,
  Vid Stojevic, Andrew~G. Green, and Simone Severini.
\newblock {Hierarchical quantum classifiers}.
\newblock {\em npj Quantum Information}, 4(1):17--19, 2018.

\bibitem{Gumpert_2017}
C~Gumpert, A~Salzburger, M~Kiehn, J~Hrdinka, and N~Calace and.
\newblock {ACTS}: from {ATLAS} software towards a common track reconstruction
  software.
\newblock {\em Journal of Physics: Conference Series}, 898:042011, oct 2017.

\bibitem{ATL-PHYS-PUB-2019-041}
{Fast Track Reconstruction for HL-LHC}.
\newblock Technical Report ATL-PHYS-PUB-2019-041, CERN, Geneva, Oct 2019.

\bibitem{refId0}
{Summers, Sioni} and {Rose, Andrew}.
\newblock Kalman filter track reconstruction on fpgas for acceleration of the
  high level trigger of the cms experiment at the hl-lhc.
\newblock {\em EPJ Web Conf.}, 214:01003, 2019.

\bibitem{HEPtrk}
{Farrell, Steven}, {Anderson, Dustin}, {Calafiura, Paolo}, {Cerati, Giuseppe},
  {Gray, Lindsey}, {Kowalkowski, Jim}, {Mudigonda, Mayur}, {Prabhat},
  {Spentzouris, Panagiotis}, {Spiropoulou, Maria}, {Tsaris, Aristeidis},
  {Vlimant, Jean-Roch}, and {Zheng, Stephan}.
\newblock The hep.trkx project: deep neural networks for hl-lhc online and
  offline tracking.
\newblock {\em EPJ Web Conf.}, 150:00003, 2017.

\bibitem{Amrouche_2019}
Sabrina Amrouche, Laurent Basara, Paolo Calafiura, Victor Estrade, Steven
  Farrell, Diogo~R. Ferreira, Liam Finnie, Nicole Finnie, Cécile Germain,
  Vladimir~Vava Gligorov, and et~al.
\newblock The tracking machine learning challenge: Accuracy phase.
\newblock {\em The Springer Series on Challenges in Machine Learning}, page
  231–264, Nov 2019.

\bibitem{farrell2018novel}
Steven Farrell, Paolo Calafiura, Mayur Mudigonda, Prabhat, Dustin Anderson,
  Jean-Roch Vlimant, Stephan Zheng, Josh Bendavid, Maria Spiropulu, Giuseppe
  Cerati, Lindsey Gray, Jim Kowalkowski, Panagiotis Spentzouris, and Aristeidis
  Tsaris.
\newblock Novel deep learning methods for track reconstruction, 2018.

\bibitem{tensorflow2015-whitepaper}
Mart\'{\i}n Abadi, Ashish Agarwal, Paul Barham, Eugene Brevdo, Zhifeng Chen,
  Craig Citro, Greg~S. Corrado, Andy Davis, Jeffrey Dean, Matthieu Devin,
  Sanjay Ghemawat, Ian Goodfellow, Andrew Harp, Geoffrey Irving, Michael Isard,
  Yangqing Jia, Rafal Jozefowicz, Lukasz Kaiser, Manjunath Kudlur, Josh
  Levenberg, Dandelion Man\'{e}, Rajat Monga, Sherry Moore, Derek Murray, Chris
  Olah, Mike Schuster, Jonathon Shlens, Benoit Steiner, Ilya Sutskever, Kunal
  Talwar, Paul Tucker, Vincent Vanhoucke, Vijay Vasudevan, Fernanda Vi\'{e}gas,
  Oriol Vinyals, Pete Warden, Martin Wattenberg, Martin Wicke, Yuan Yu, and
  Xiaoqiang Zheng.
\newblock {TensorFlow}: Large-scale machine learning on heterogeneous systems,
  2015.
\newblock Software available from tensorflow.org.

\bibitem{Havlcek2019SupervisedLW}
Vojtech Havl{\'i}cek, Antonio~D. C{\'o}rcoles, Kristan Temme, Aram~W. Harrow,
  Abhinav Kandala, Jerry~M. Chow, and Jay~M. Gambetta.
\newblock Supervised learning with quantum-enhanced feature spaces.
\newblock {\em Nature}, 567:209--212, 2019.

\bibitem{COBYLA}
M.~J.~D. Powell.
\newblock A direct search optimization method that models the objective and
  constraint functions by linear interpolation.
\newblock {\em Mathematics and Its Applications}, 275, 1994.

\bibitem{SPSA1997}
James C.Spall.
\newblock A one-measurement form of simultaneous perturbation stochastic
  approximation.
\newblock {\em Automatica}, 33, 1997.

\bibitem{SPSA2020}
James C.Spall.
\newblock Adaptive stochastic approximation by the simultaneous perturbation
  method.
\newblock {\em IEEE Transaction on Automatic Control}, 45, 2000.

\bibitem{Mitarai_2018}
K.~Mitarai, M.~Negoro, M.~Kitagawa, and K.~Fujii.
\newblock Quantum circuit learning.
\newblock {\em Physical Review A}, 98(3), Sep 2018.

\bibitem{Qulacs}
{Qulacs}, 2018.

\bibitem{Dua:2019}
Dheeru Dua and Casey Graff.
\newblock {UCI} machine learning repository, 2017.

\bibitem{Gaoeaat9004}
X.~Gao, Z.-Y. Zhang, and L.-M. Duan.
\newblock A quantum machine learning algorithm based on generative models.
\newblock {\em Science Advances}, 4(12), 2018.

\bibitem{classicalsvm}
Bernhard~E. Boser, Isabelle~M. Guyon, and Vladimir~N. Vapnik.
\newblock A training algorithm for optimal margin classifiers.
\newblock {\em COLT 92: Proceedings of the fifth annual workshop on
  Computational learning theory}, page 144, 1992.

\bibitem{aaronson15}
Scott Aaronson.
\newblock Read the fine print.
\newblock {\em Nature Physics}, 11(4):291--293, 2015.

\bibitem{tang2019quantum}
Ewin Tang.
\newblock A quantum-inspired classical algorithm for recommendation systems.
\newblock In {\em Proceedings of the 51st Annual ACM SIGACT Symposium on Theory
  of Computing}, pages 217--228, 2019.

\bibitem{arrazola2019quantum}
Juan~Miguel Arrazola, Alain Delgado, Bhaskar~Roy Bardhan, and Seth Lloyd.
\newblock Quantum-inspired algorithms in practice.
\newblock {\em arXiv preprint arXiv:1905.10415}, 2019.

\bibitem{benedetti16a}
Marcello Benedetti, John Realpe-G{\'o}mez, Rupak Biswas, and Alejandro
  Perdomo-Ortiz.
\newblock Estimation of effective temperatures in quantum annealers for
  sampling applications: {A} case study with possible applications in deep
  learning.
\newblock {\em Physical Review A}, 94(2):022308, 2016.

\bibitem{booth2017partitioning}
M~Booth, SP~Reinhardt, and A~Roy.
\newblock Partitioning optimization problems for hybrid classcal/quantum
  execution, 2017.

\bibitem{digitalannealer}
Fujitsu.
\newblock Digital annealer, 2020.
\newblock
  https://www.fujitsu.com/global/services/business-services/digital-annealer/.

\bibitem{havlivcek2019supervised}
Vojt{\v{e}}ch Havl{\'\i}{\v{c}}ek, Antonio~D C{\'o}rcoles, Kristan Temme,
  Aram~W Harrow, Abhinav Kandala, Jerry~M Chow, and Jay~M Gambetta.
\newblock Supervised learning with quantum-enhanced feature spaces.
\newblock {\em Nature}, 567(7747):209, 2019.

\bibitem{schuld2017implementing}
M~Schuld, M~Fingerhuth, and F~Petruccione.
\newblock Implementing a distance-based classifier with a quantum interference
  circuit.
\newblock {\em EPL (Europhysics Letters)}, 119:60002, 2017.

\bibitem{mckiernan2019automated}
Keri~A. McKiernan, Erik Davis, M.~Sohaib Alam, and Chad Rigetti.
\newblock Automated quantum programming via reinforcement learning for
  combinatorial optimization, 2019.

\bibitem{temme2017error}
Kristan Temme, Sergey Bravyi, and Jay~M Gambetta.
\newblock Error mitigation for short-depth quantum circuits.
\newblock {\em Physical review letters}, 119(18):180509, 2017.

\bibitem{analyticgradient}
Maria Schuld, Ville Bergholm, Christian Gogolin, Josh Izaac, and Nathan
  Killoran.
\newblock Evaluating analytic gradients on quantum hardware.
\newblock {\em Phys. Rev. A}, 99(032331), 2019.

\bibitem{unsupervised-quantum-data}
Gael Sent{\'\i}s, Alex Monr{\`a}s, Ramon Mu{\~n}oz-Tapia, John Calsamiglia, and
  Emilio Bagan.
\newblock Unsupervised classification of quantum data.
\newblock {\em Physical Review X}, 9(4):041029, 2019.

\bibitem{no-free-lunch}
Kyle Poland, Kerstin Beer, and Tobias~J Osborne.
\newblock No free lunch for quantum machine learning.
\newblock {\em arXiv preprint arXiv:2003.14103}, 2020.

\bibitem{qcnn}
Iris Cong, Soonwon Choi, and Mikhail~D Lukin.
\newblock Quantum convolutional neural networks.
\newblock {\em Nature Physics}, 15(12):1273--1278, 2019.

\bibitem{quantum-anomaly-detection}
Nana Liu and Patrick Rebentrost.
\newblock Quantum machine learning for quantum anomaly detection.
\newblock {\em Physical Review A}, 97(4):042315, 2018.

\bibitem{learning-purity}
Lukasz Cincio, Yi{\u{g}}it Suba{\c{s}}{\i}, Andrew~T Sornborger, and Patrick~J
  Coles.
\newblock Learning the quantum algorithm for state overlap.
\newblock {\em New Journal of Physics}, 20(11):113022, 2018.

\bibitem{quantum-fidelity-estimation}
Marco Cerezo, Alexander Poremba, Lukasz Cincio, and Patrick~J Coles.
\newblock Variational quantum fidelity estimation.
\newblock {\em Quantum}, 4:248, 2020.

\bibitem{classification-quantum-data}
Farzad Shahi and Ali~T Rezakhani.
\newblock Binary classification of quantum states: Supervised and unsupervised
  learning.
\newblock {\em arXiv preprint arXiv:1704.01965}, 2017.

\bibitem{learning-coherent-states}
Gael Sent{\'\i}s, M{\u{a}}d{\u{a}}lin Gu{\c{t}}{\u{a}}, and Gerardo Adesso.
\newblock Quantum learning of coherent states.
\newblock {\em EPJ Quantum Technology}, 2(1):1--22, 2015.

\bibitem{doi:10.1063/PT.3.4227}
Karl van Bibber, Konrad Lehnert, and Aaron Chou.
\newblock Putting the squeeze on axions.
\newblock {\em Physics Today}, 72(6):48--55, 2019.

\bibitem{lloyd1996universal}
Seth Lloyd.
\newblock Universal quantum simulators.
\newblock {\em Science}, pages 1073--1078, 1996.

\bibitem{RevModPhys.86.153}
I.~M. Georgescu, S.~Ashhab, and Franco Nori.
\newblock Quantum simulation.
\newblock {\em Rev. Mod. Phys.}, 86:153--185, Mar 2014.

\end{thebibliography}
\bibliographystyle{unsrt}

\end{document}